\documentclass[useAMS]{mn2e}
\usepackage{graphicx}
\usepackage{epstopdf}
\usepackage{color}

\title[Cool core clusters \& SZ effect]{Signature of cool core in SZ clusters: a multiwavelength approach}

\author[Pipino \& Pierpaoli]{A. Pipino\thanks{pipino@usc.edu,pipino@astro.ucla.edu}$^{1,2,3}$, \& E. Pierpaoli$^1$\\
$^1$Department of Physics \& Astronomy, University of Southern California, Los Angeles 90089-0740, USA\\
$^2$Dipartimento di Fisica, Sez. Astronomia, Universita'di Trieste, via GB Tiepolo 11, Trieste 34100, Italy\\
$^3$Department of Physics \& Astronomy, University of California, Los Angeles, 430 Portola Plaza, Los Angeles 90095, USA}

\long\def\symbolfootnote[#1]#2{\begingroup%
\def\thefootnote{\fnsymbol{footnote}}\footnote[#1]{#2}\endgroup}

\begin{document}

\date{Accepted 2010 January 19.  Received 2009 December 4; in original form 2009 August 10}

\pagerange{\pageref{firstpage}--\pageref{lastpage}} \pubyear{2008}

\maketitle

\label{firstpage}

\begin{abstract}
We use the high quality pressure profiles of 239 galaxy clusters made
available by the ACCEPT project (Cavagnolo et al. 2009) in order
to derive the expected Sunyaev Zeldovich (SZ) signal in a variety of cases
that hardly find a counterpart in the simulations.
We made use of the Melin et al. (2006) cluster selection function for both
the South Pole Telescope (SPT) and Planck instruments.
Prior knowledge of the entropy profiles of the same clusters
allows us to study the impact of cool cores (CC) in cluster detection
via SZ experiment and to test if this introduces a bias
in inferred quantities as, e.g., the mass function.

We infer a clear effect of the CC on the central Compton parameter $y_0$.
We thus validate the suggestions by McCarthy et al. (2003), namely that at a given mass clusters
with higher entropy levels show a lower $y_0$ than their low
entropy counterparts, on a much
larger sample of clusters.
For a high resolution experiment like SPT, we expect that
the fraction of detected clusters with respect to the total
to decline at masses around $\sim 2\times10^{14} M_{\odot}$.
For Planck this happens at a somewhat higher mass.
We find that the presence of CCs introduces a small bias in cluster
detection, especially around the mass at which the 
performance of the survey begins to decrease.
If the CC were removed, a lower overall fraction
of detected clusters would be expected.
In order to estimate the presence of such a bias by means
of SZ only surveys, we show that the ratio between $y_0$ and $y_{int}$
anti-correlates with the cluster central cooling time.
If multi-band optical cluster surveys are either available for a cross-match
or a follow-up is planned, we suggest that likely CC clusters
are those with a Brightest Cluster Galaxy (BCG) at least 0.3 magnitudes bluer than the average.
A more robust estimate of the CC presence is given by
UV-optical colours of the BCG, like the NUV-r, whose values can be 4 magnitudes
off the NUV-r equivalent of the red sequence, in clusters with low
excess entropy. We also find correlation of the $y_0/y_{int}$ ratio
with H$\alpha$, IR and radio luminosities.
We argue that the analysis of a combined SZ/optical/UV surveys can be also used
to shed light on the suggested CC evolution with redshift.

\end{abstract}


\section{Introduction}

There are great expectations for Sunyaev-Zeldovich effect (Sunyaev \& Zeldovich, 1970, Carlstrom et al., 2002, SZ hereafter) based
clusters surveys. The ones already started (the South Pole Telescope, SPT, Ruhl et al., 2004, the Atacama Cosmology Telescope, ACT) or  up-coming (Planck)
will contribute to a dramatic increase in the number of galaxy clusters available
for applications ranging from the hot intracluster medium physics to cosmological parameter determination (Barbosa et al. 1996, Benson et al., 2002, Weller \& Battye, 2003).

Preliminary works explored the selection functions of different instruments
as a function of instrumental characteristics (including noise) and of the adopted technique
to actually detect clusters in the SZ signal.
Several filtering methods like the the matched filter approach (Melin
et al., 2006), multi-frequency Wiener filtering (Pierpaoli et al., 2005) and others (e.g. Vale \& White, 2006, Pires et al. 2006, Diego et al., 2002, Schaefer
et al. 2006) have been studied
in relation to one or more SZ experiments.
We refer to Melin et al. (2006), Pierpaoli et al. (2005), for a comprehensive discussion of (differing)
cluster detection techniques and Leach et al. (2009) for the latest comparison of their performances
on simulated Planck SZ maps.
Most of these works adopted either analytic models or the results of N-body simulations (in which the gas follows the Dark Matter
particles, e.g. Vale \& White, 2006) to mimic the SZ signal from the galaxy clusters.
Indeed, these works were useful to better assess the capabilities of each instrument and plan its use.
However, now that the first SZ detected clusters (Staniszewski et al, 2008) are available it is important
to pay more attention to the intracluster medium (ICM) details.
The fact that relations between cluster physical properties deviate from a self-similar scaling (Allen \& Fabian, 1998, Borgani et al., 2002)
reminds us that the treatment of the gas physics must be very accurate.
For instance, McCarthy et al. (2003a,b) showed that the presence of cool cores\footnote{See  
McCarthy et al. 
2004; 2008 for a detailed discussion of the diversity in the cluster population, including the distinction between 
cool core versus non-cool core clusters.} (CC, hereafter)
in cluster of galaxies may affect the SZ signal. 
McCarthy et al. studied the case in which the \emph{pre-heating} is the cause of
the high entropy floor seen in non CC (NCC) clusters of galaxies and showed that the temperature of the gas near the centre of the cluster
increases. At the same time, the density decreases. Since the effect is larger
on the density rather than the temperature, the net result is that
the gas pressure in the central regions of the clusters decreases.
The SZ Compton parameter is the integral on the line of sight
of the pressure profile (Eq.~\ref{SZE}), hence we expect NCC clusters,
namely objects in which the excess entropy is larger, to have a shallower
pressure profile (lower SZ signal), than CC clusters of the same mass.

While differences in the SZ vs X-ray properties between CC and NCC clusters
have been emphasized also by observational works (e.g. Morandi et al., 2007),
previous estimates of both the cluster detection threshold and the completeness of SZ surveys, in terms of either
the cluster mass or their SZ integrated signal, 
made use of mock cluster maps, did not pay
attention to the presence of CCs or considered it negligible (Schafer \& Bartelmann, 2007).
In particular, the emission from clusters
was often taken to trace the Dark Matter potential, without
a detailed modelling of the ICM physics. 
The goal of the present paper is to be a first step in filling this gap.

Our approach is rather different in another aspect. In fact, we use the pressure profiles for 239 clusters
obtained from high-quality Chandra data.
We then recover their (expected) SZ signal by a direct integration of their pressure profiles and
use it to infer which properties of the cluster (if any) make
it detectable in a SZ survey.
Our main aim is to test McCarthy et al.'s hypothesis on a more
general (and larger) sample of clusters with respect to McCarthy et al. (2003b). We then focus on
the bias on the recovered clusters mass and abundance
in present on-going cluster surveys induced by the presence of CC.
In particular we will ask ourselves the questions:
are CC clusters more likely to be detected than NCC clusters of the same mass? What happens
for clusters around the instrument limit\footnote{The mass at which the
detection rate falls below the 90\%. This is set by instrument properties, the noise,
but also by the detection method.} mass
for detection? At a given mass, is it easier for CC clusters
to make such a threshold than NCC clusters? If so, should
we expect a bias in the mass functions of the clusters derived
by SZ-only surveys? 
Our sample is not complete in any sense. Therefore we
cannot forecast, e.g., the cluster mass function that can be
constructed from SZ-only surveys.
However, our clusters cover a large range in masses, redshift, entropy profiles.
In practice we complement the modelling approach (Vale \& White, 2006, Shaefer \& Bartelmann, 2007,
Nagai, 2006, Nagai et al., 2007) presenting
results that sample a large variety of conditions of the hot intracluster medium, with
a spread that is fair representation of what happens in the local Universe.
In this sense the goal is to make a first quantitative estimate
of the CC bias around the instrument limit mass. The presence of the bias and its estimate are linked to the chosen
method for cluster detection is SZ maps; however our results can be read as
a warning and possibly extended to other cluster selection criteria.

We further explore the connection between the presence of CC and the properties
of the Brightest Central Galaxies (BCGs). Indeed, recent studies in several bands have reported
examples of ongoing star formation in the BCGs (Cardiel et al. 1998, Crawford et al. 1999, Edge 2001, 
Goto 2005, McNamara et al. 2006, { O'Dea et al. 2008}, Bildfell et al. 2008, Cavagnolo et al. 2008, 
Rafferty et al. 2008, Edwards et al. 2008, Pipino et al., 2009, Sanderson et al., 2009); most of these BCGs reside in CC
clusters and are located very close to the peak of the X-ray emission. 
In particular, it seems that up to the 25\% of the BCGs are
somehow active. Therefore, another goal of this paper is to extended
the body of evidence for the CC-BCG connection. In particular, we will investigate if a correlation between star formation
indicators like optical blue cores, UV excess, H$\alpha$ emission, high IR and radio luminosities
and the SZ effect from CC clusters exists. We will
suggest a multiwavelenght approach in order to complement
SZ-only based surveys to assess the presence of a bias
induced by CCs. The same technique can be applied
to other science cases, as, e.g., the study of the evolution
of the CC fraction in a larger sample of clusters than
those provided by X-ray data only.

We will present our results for either a high resolution
experiment (SPT) and a lower resolution one (Planck).

The plan of the paper is the following. The cluster sample
and the SZ signal that we expect are characterized in Secs.~\ref{sample}
and ~\ref{signal}, respectively. We will present our results
in Sec.~\ref{res}, discuss their implications in Sec.~\ref{disc} and draw our conclusions in Sec.~\ref{conc}.

\section{Modelling the SZ signal.}

\subsection{The cluster sample}
\label{sample}

\begin{figure}
\begin{center}
\includegraphics[width=3in,height=3in]{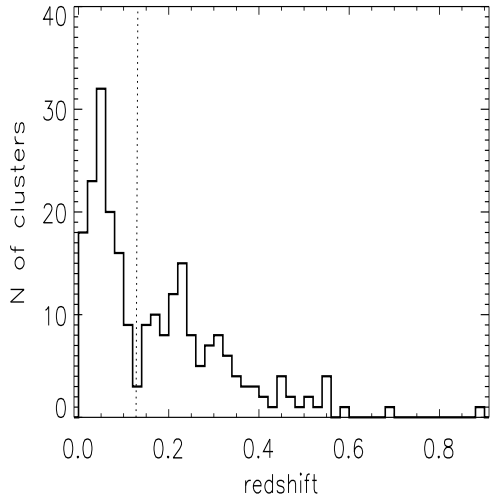}
\caption{Redshift distribution of the adopted sample of galaxy clusters. The dotted
vertical line marks the median redshift ($\sim$0.13) of the distribution.}
\label{accept_hist1}
\end{center}
\end{figure}

We make use of the publicly available data made possible by the ACCEPT project (Cavagnolo et al. 2009).
The catalog comprises of 239 galaxy clusters with accurate temperature, density, entropy and pressure
profiles reduced in a homogeneous way from public Chandra data. The catalog covers
the temperature range 1-20 keV with redshifts ranging from 0.05 to 0.89 (see Fig.~\ref{accept_hist1}).
We refer to Cavagnolo et al. (2009,
and references therein) for details on the data reduction and further catalog specifics.
Here we note that this catalog is neither flux limited, nor volume limited. However, since ACCEPT clusters come from a large number
of observing programs, it is quite unlikely that it is biased toward a particular
class of clusters. Moreover, as noted by Cavagnolo et al., almost all (94\%) of the HIFLUGCS (Reiprich \& Bohringer, 2002)
clusters are in ACCEPT. Therefore we will use HIFLUGCS clusters available in ACCEPT as a flux-limited sub-sample
in order to make more quantitative estimates in the following.

\begin{figure}
\begin{center}
\includegraphics[width=3.5in,height=4in]{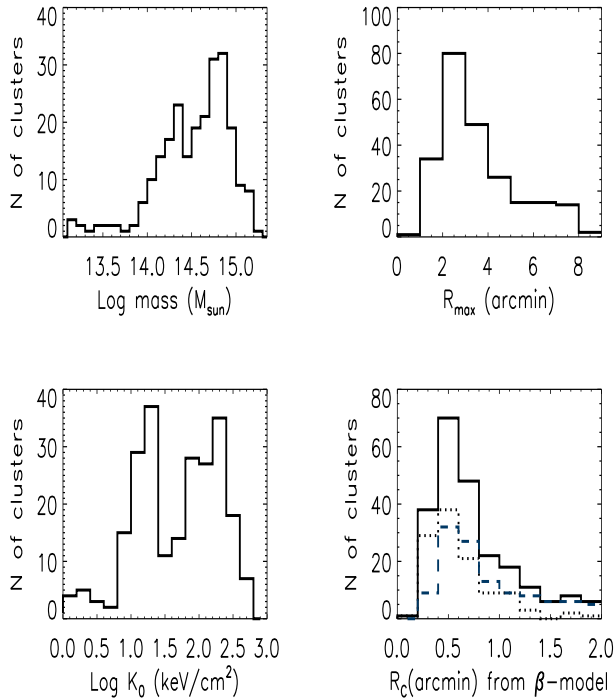}
\caption{Properties of the adopted sample of galaxy clusters. Top-left: M$_{500}$ distribution. Bottom-left:
entropy distribution. Top-right: Maximum radius probed by X-ray observation, namely the radius
of the largest annulus available from ACCEPT for a direct integration of the pressure profile (projected in the sky). Bottom-right: 
Core radius distribution inferred from a $\beta$-model fitting (see text). In this panel, CC 
clusters are displayed with a dotted-line, NCC with a dashed 
line.}
\label{accept_hist}
\end{center}
\end{figure}

We estimate the mass of our clusters (not provided in ACCEPT) as 
\begin{equation}
M_{500}=3.02 \times (kT_X/5keV)^{1.53} H_0/H(z) h^{-1} 10^{14} M_{\odot}
\end{equation}
given by Vikhlinin et al. (2009).
This relation will set the baseline for our discussion. 
We note that our choice does not affect the results of the calculations
shown in remainder of the paper. It is needed only to infer a mass
for the objects we are dealing with.
The mass distribution of our cluster sample is shown in the top-left panel
of Fig.~\ref{accept_hist}.

The radius of the clusters ($R_{\Delta_c}$) were calculated by Cavagnolo following
Arnaud et al. (2002):
\begin{eqnarray}
R_{\Delta_c} &=& 2.71 \mathrm{~Mpc~}
\beta_T^{1/2}
\Delta_{\mathrm{z}}^{-1/2}
(1+z)^{-3/2}
\left(\frac{kT_X}{10 \mathrm{~keV}}\right)^{1/2}\\
\Delta_z &=& \frac{\Delta_c \Omega_M}{18\pi^2\Omega_z} \nonumber \\
\Omega_z &=& \frac{\Omega_M (1+z)^3}{[\Omega_M
(1+z)^3]+[(1-\Omega_M-\Omega_{\Lambda})(1+z)^2]+\Omega_{\Lambda}} \nonumber
\end{eqnarray}
where $R_{\Delta_c}$ is in units of $h_{70}^{-1}$, $\Delta_c$ is
the assumed density contrast of the cluster at $R_{\Delta_c}$, and
$\beta_T$ is a numerically determined
 normalization for the virial relation.
The maximum radius ($R_{max}$) probed by X-ray observations is typically 1/2 of
the virial radius (Fig.~\ref{accept_hist}, the top-right panel).  
Since clusters are at different redshifts and their projected angle $\theta_{max}$ on the sky 
matters, we converted the radii into angles after having calculated the distance of
the clusters in a flat $\Lambda$CDM Universe with $\Omega_m=0.3$, $\Omega_{\Lambda}=0.7$ and
$H_0= 70 \rm km\, s^{-1}\, Mpc^{-1}$ (these are the value adopted in Cavagnolo et al., 2009 - we
use them in order to be consistent with their estimates of, e.g., luminosities). 
All clusters have properties measured out to (at least) 1 arcmin. Therefore
all of them could in principle be resolved by an instrument like SPT (Bartlett, 2006). Only few of them,
instead, have dimension larger than 5 arcmin (namely the FWHM resolution of Planck
in the 217 GHz channel).

We also fitted a single $\beta$-model (Cavaliere \& Fusco-Femiano, 1976) to our clusters
in order to estimate the cluster core radius that we will employ to determine if 
a cluster can be detected (see Section~\ref{melin}). In order to be consistent
with the assumption behind theoretical detection limit that we will use, we fixed $\beta=0.66$.
We did not attempt to fit double $\beta$-models in CC clusters
where it seems to be required (Cavagnolo et al. , 2009).
Instead, in order to minimize the effect of the CC,
we avoided the very central regions ($log R/R_{200} < -2$) of the clusters.
In practice, the SZ signal obtained by the integration of the $\beta$ model
will give an estimate of the signal obtained for real clusters
when the CC-effect is removed.
The core radius ($R_c$) distribution of our cluster sample is shown in the bottom-right panel
of Fig.~\ref{accept_hist}. Again, we show the angle on the sky rather than the radius in physical units. 
Most of the clusters have core size around 0.5 arcmin.
Therefore, they must be treated as extended sources in order
to properly evaluate the possibility to be detected by a given instrument.
We note that the values for $R_c$ in CC clusters are marginally smaller than
those for NCC. Such a trend is visible in the distribution of clusters
core angles $\theta_c$ (the bottom-right panel of Fig.~\ref{accept_hist}). 
This is a secondary bias linked to the presence of the CC, the primary
being the increased Compton parameter, that the reader should be aware of.
{ Note that the above discussion holds for the actual ACCEPT redshift distribution
and will be used in the rest of the paper unless otherwise stated. As a matter of fact, w
e will also present a case (see Sec. 3.2) in which all the clusters are placed
to higher redshift, while preserving their mass, to check our results in the case
of a more typical redshift distribution. In that case, the resulting $\theta_c$
will be smaller - and with a different distribution - than the ones shown
in Fig.~\ref{accept_hist}}.

From the bottom-left panel of Fig.~\ref{accept_hist}, we note that
there is a clear bimodality in the cluster entropy distribution
defined as the \emph{entropy floor} $K_0$ in units of  $\rm keV cm^2$.
Cavagnolo et al. (2009, see also Rafferty et al., 2008) claim that this bimodality is not induced
by a bias in the cluster selection.
We have also to keep in mind that the fraction of CC clusters in ACCEPT is $\sim 40\%$, somewhat lower
than the typical $\sim 50-60\%$ reported in the literature (Peres et al., 1998), and that the ACCEPT lacks clusters with $K_0$ below 10 $\rm keV cm^2$
at z$>0.1$. We expect the latter to be less important. In fact, entropy levels below 10 $\rm keV cm^2$
are typical of galaxy groups, namely systems not observed in SZ due to their very low signal.
The distribution of clusters as a function of the central cooling time follows
the one as a function of $K_0$ (see Cavagnolo et al. 2009).
In the following we refer to CC clusters as those with entropy $K_0$ below 25 $\rm keV cm^2$
or cooling time below 1 Gyr, being the differences between the two classification
negligible in our results.

\subsection{SZ signal from our clusters}
\label{signal}

In
the following we will express the SZ effect through the Compton $y$ parameter
given by the following equation:
\begin{equation}
y(\theta)= \int \sigma_T n_e \frac{k_B T_e}{m_e c^2} dl
\label{SZE}
\end{equation}
where $\sigma_T$ is the
Thomson cross section, $k_B$ is the Boltzmann constant, $c$ is the
speed of light in vacuum, $m_e$ is the electron mass, $T_e$ the
electron temperature, and the integration is along the line of sight.
The central Compton parameter is $y_0=y(\theta=0)$, whereas
we recall that the integral effect within a given angle $\theta$ is:
\begin{equation}
y_{int}(<\theta)= 2\pi \int_0^{\theta} y(\theta ') \theta ' d\theta '
\label{SZE1}
\end{equation}

\begin{figure}
\begin{center}
\includegraphics[width=3in,height=2.5in]{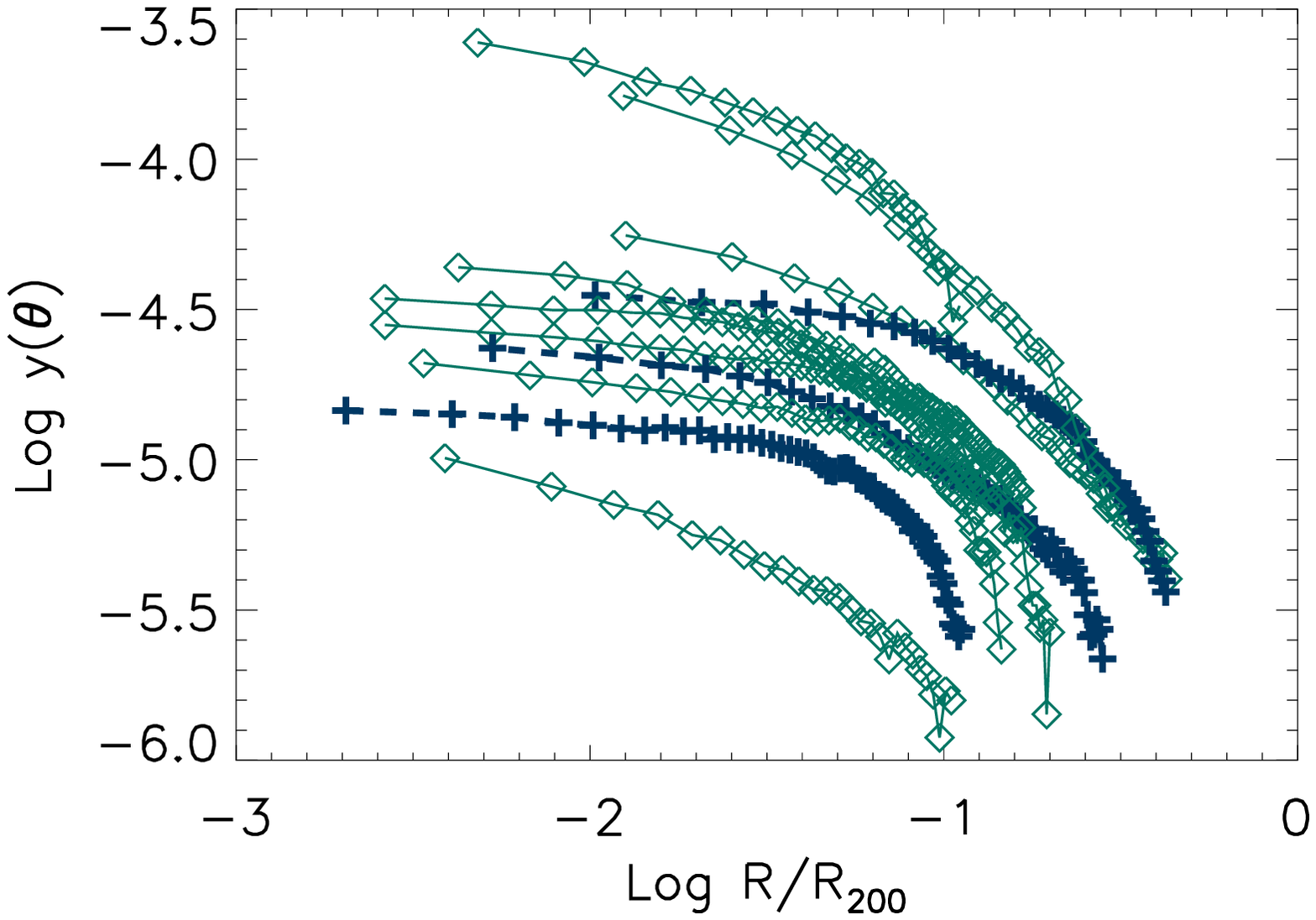}
\includegraphics[width=3in,height=2.5in]{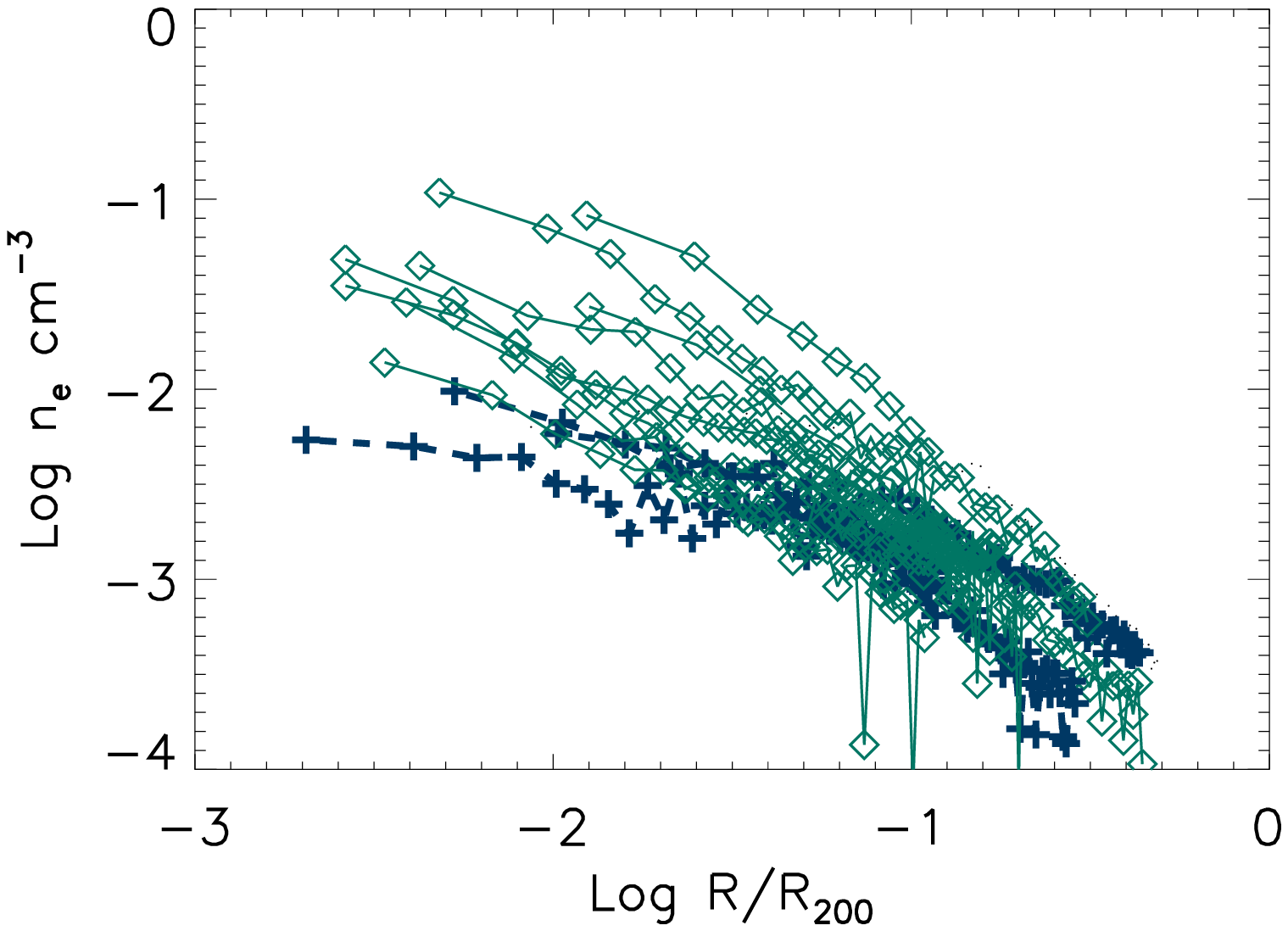}
\includegraphics[width=3in,height=2.5in]{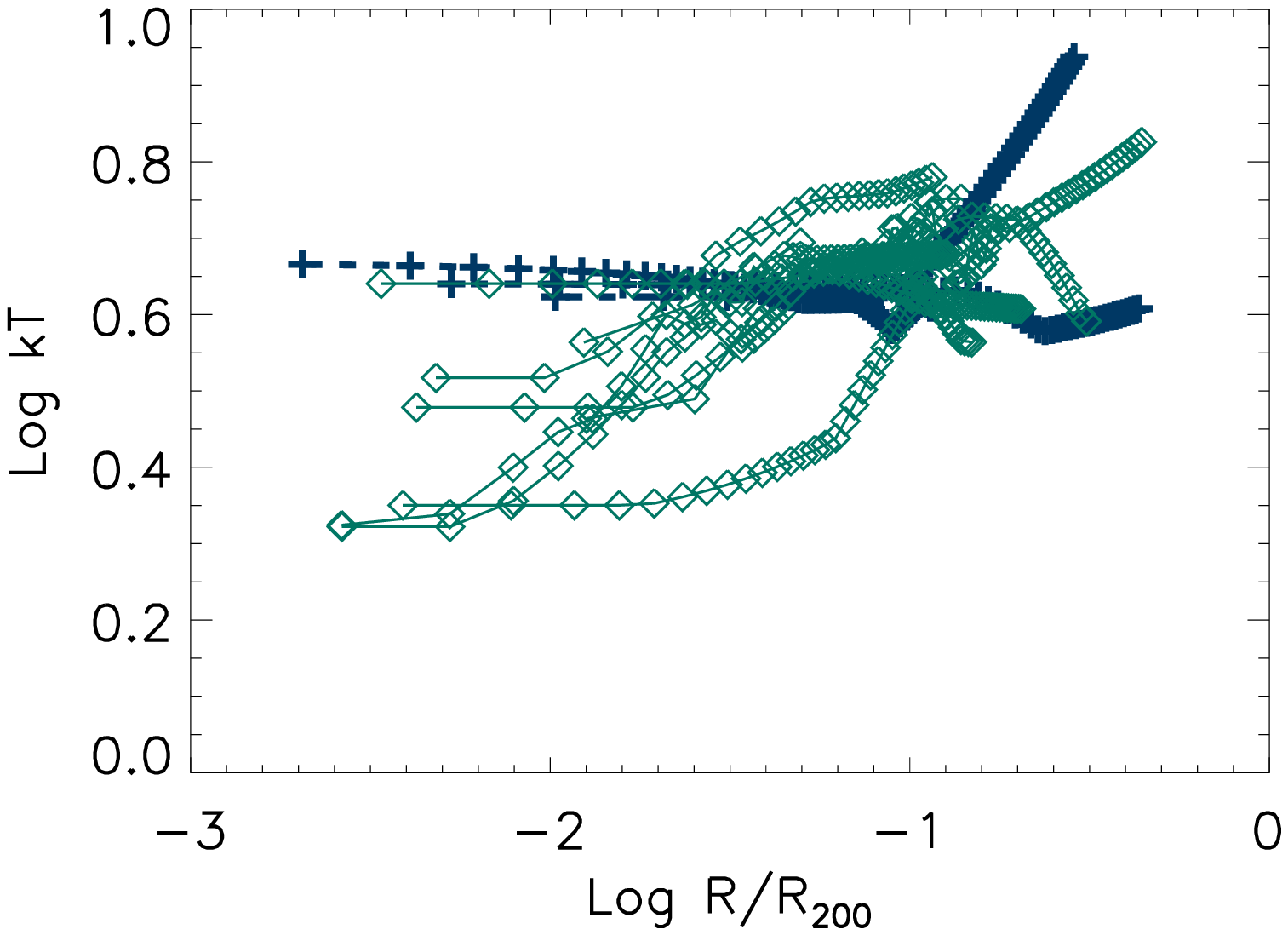}
\caption{Compton parameter (upper panel), density (central panel) and
temperature (lower panel) radial profiles for clusters
with mass $\sim 2.5\times10^{14} M_{\odot}$. CC clusters are presented
as diamonds joined by solid lines, whereas NCC clusters as crosses
joined by dashed lines. The CC presence enhances the central electron
density by an order of magnitude with respect to the NCC case, whereas
the temperature drops by a factor of 2, hence the steepening of
the y$_{int}(\theta)$ in CC clusters.}
\label{prof}
\end{center}
\end{figure}

We obtain the Compton parameter $y$ by direct integration of the pressure
profiles provided by ACCEPT. In Fig.~\ref{prof}
we show some examples of the Compton parameter radial profiles derived by us compared
to the input temperature and density profiles.
 For the sake of clarity, we
focus on clusters
with mass $\sim 2.5\times10^{14} M_{\odot}$, as we will see in the following that this is
roughly the mass at which the CC-induced bias is more relevant.
Clearly, CC clusters exhibit steeper profiles in $y(\theta)$
than NCC clusters, due to the fact the CC, while decreasing the central temperature
by a only factor of a few, enormously enhances the central density.
The pressure profile is thus steeper in CC clusters. However, in order to get $y(\theta)$
we have to further integrate the pressure profile along the line of sight (Eq.~\ref{SZE}),
leading to Compton parameter profiles less steep than the density profiles
from which they arose.
Note that the ACCEPT pressure profiles allow us a direct estimate of the SZ effect
out to $R_{max}$.

Although $R_{max} \sim 0.5\times R_{200}$ (Cavagnolo et al. 2009),
this radius is not constant for all clusters. Moreover, clusters
have differing redshifts, therefore they will cover differing angles in the sky
(see top-right panel of Fig.~\ref{accept_hist}). 

\begin{figure}
\includegraphics[width=3in]{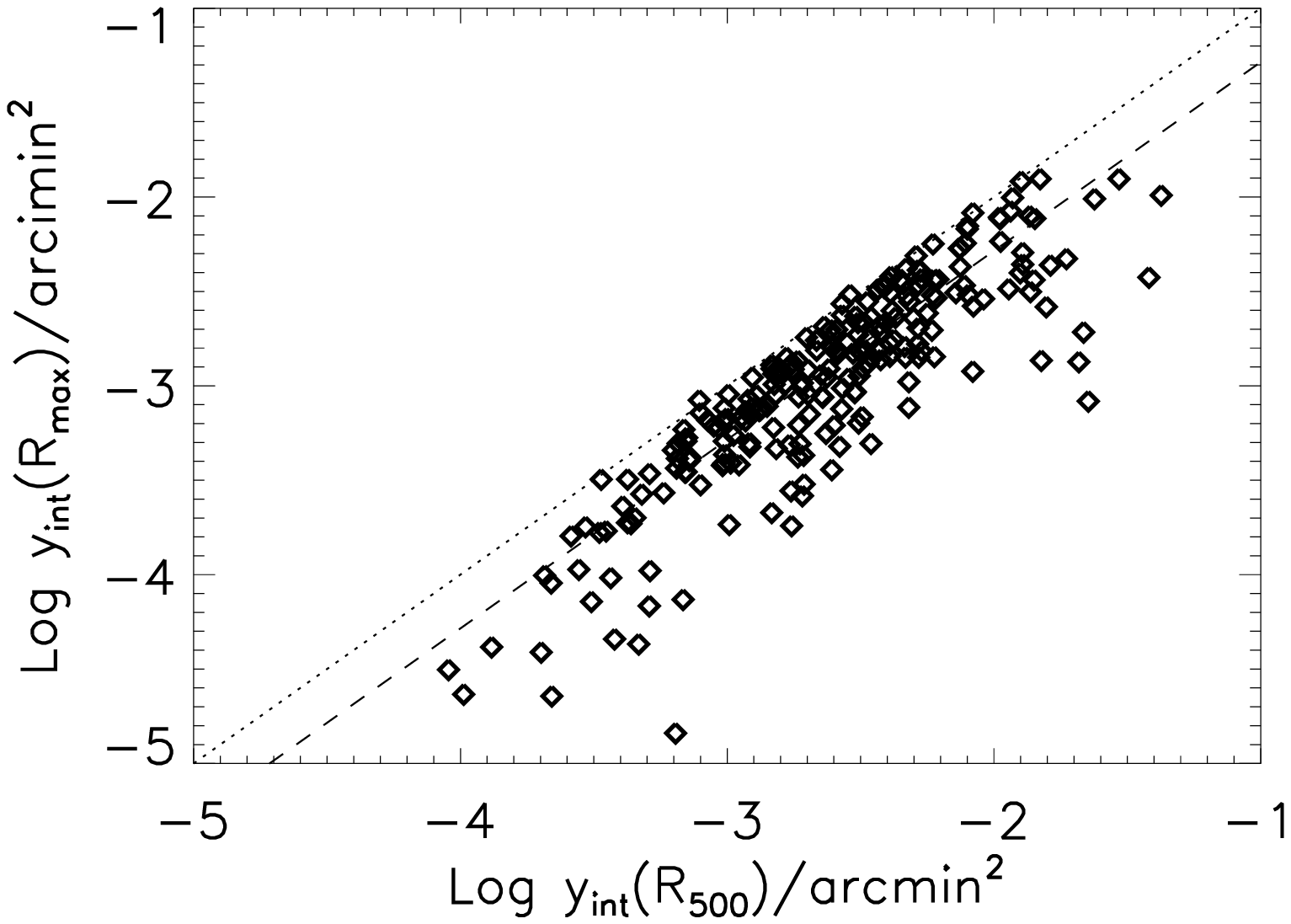}
\includegraphics[width=3in]{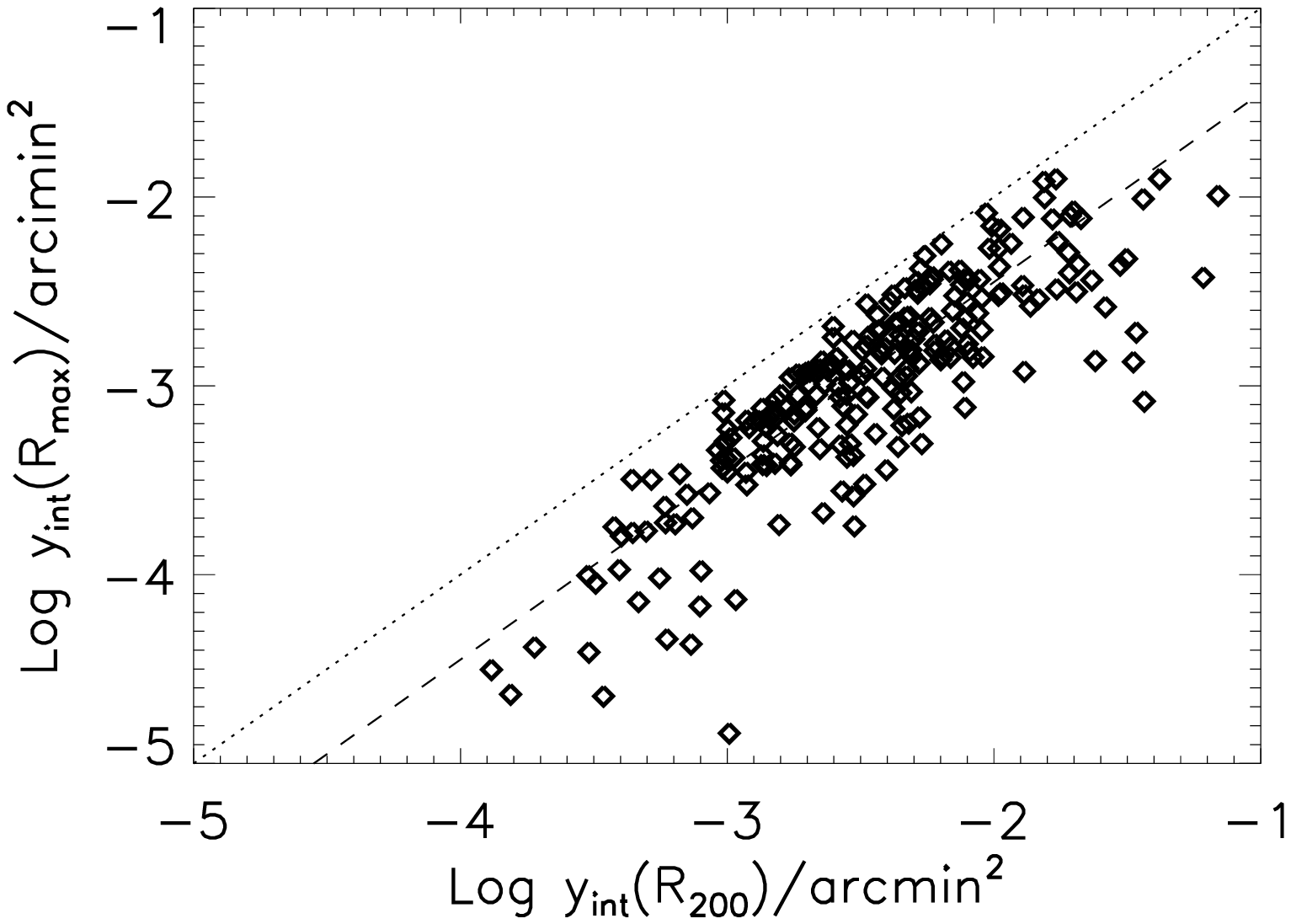}
\caption{Comparison between the integrated Compton parameter at $R_{max}$ 
(i.e. obtained by direct integration of the profiles coming from X-ray data
and the extrapolated values at $R_{500}$ and $R_{200}$ (see text for details).
The dotted line is the 1:1 relation to guide the eye. The dashed line gives the median
offset from the 1:1 relation.}
\label{max_vs}
\end{figure}

We will present also results out to $R_{500}$ ($R_{200}$) by adding the SZ signal at radii $R_{max} < R< R_{500}$
($R_{200}$)
estimated by the beta-model fitted profile to the value for $y_{int}(<R_{max})$
obtained from direct integration of the ACCEPT pressure profiles.
In this way we minimize the impact the $\beta$-model fitting procedure and maintain
the effect of the CC.
The comparison between these quantities is given in Fig.~\ref{max_vs}. In this figure, 
the dotted line is the 1:1 relation to guide the eye. The dashed line give the median
offset from the 1:1 relation. In particular, the median offset between $y_{int}(<R_{max})$
and $y_{int}(<R_{500})$ is less than a factor of 2, since for many clusters
$R_{max}\sim R_{500}$. In the case of the extrapolation of the Compton parameter
out to $R_{200}$, the median difference is a factor of 2.8.

As for the SZ signal obtained through the integration of the $\beta$
model, we integrate  out to both $R_{max}$ and $R_{500}$ the latter case
being similar to the assumptions of Melin et al. (2006), who
integrated out to 10$R_{c}$ (see below).

A comparison between $y_{int}(<R_{max})$ calculated in the above mentioned way and the
SZ calculated from the beta-model alone out to the $R_{max}$,
will then highlight the role of CC in boosting the SZ signal.

\subsection{Detection limit}
\label{melin}

The detection of a cluster in a SZ map not only depends on its flux,
but also on its extension on the sky and the characteristics of competing
signals. These quantities, then, should
be coupled with the filtering method chosen to locate the cluster
in a SZ map and  for reconstruction of the signal.
The selection function will be a product of all these factors.
In this work we make use of the limits calculated by Melin et al. (2006)
by means of their multi-frequency matched filter applied on Monte-Carlo simulated maps.

It suffices to say that Melin et al. assumed an isothermal $\beta=0.66$ model profile 
out to 10$\theta_{c}$ as the filter template for the SZ signal, where
only the core radius $\theta_c$ is free to vary, and tested the performances
of several instruments.
Here we make use of their results (c.f. their Fig.2) to quantify the fraction of our
clusters that exceed a given sensitivity limit. In particular, Melin et al. (2006)
show their filter noise in terms of integrated SZ flux as a function of the
cluster core radius. 
The curves in their Fig.2, thus, show the minimum signal that a cluster
must emit in order to achieve a S/N=1, given its core radius, a chosen
instrument and, of course, using the filtering scheme proposed by Melin et al. (2006).
In the following we will always refer to the signal needed to attain
S/N=5. This is a conservative estimate of sensitivity limit for a detection
of a given instrument.
We note that lowering the S/N threshold to 1 will consequently increase
the number of clusters that can be detected and reduce the minimum
cluster mass that an instrument can probe.
In real life, instead, things are much more complicated 
and other factors can decrease the performances of a instrument. For the Planck
case, Leach et al (2009) note that many of the recovered clusters are to be considered
resolved, and thus emit on scales where the contamination from CMB is not negligible. Moreover,
small scale Galactic emission and the background of extra-galactic sources further
complicate the detection when included in the simulations.
Therefore, it seems likely that previous estimates of the detection
threshold (including Melin et al.'s one that we adopt) should be revised upward by
a factor 3-5 (Melin et al., priv.comm).
It is important to note that, for a survey like Planck, the sensitivity is much less affected by the angular
extension of the sky than for the high resolution case (SPT).
{ The ACCEPT cluster distribution is not representative of the number counts of clusters above
a given mass (see e.g. Hallman et al., 2007). An accurate study should take into account
that a larger number of relatively massive clusters is expected at redshifts
higher than the typical ACCEPT redshift. In Sec. 3.2 we will present a case
in which we randomly assign the ACCEPT clusters a higher redshift in order
to assess this issue. 
\clearpage
\begin{figure*}
\includegraphics[width=\textwidth,height=6in]{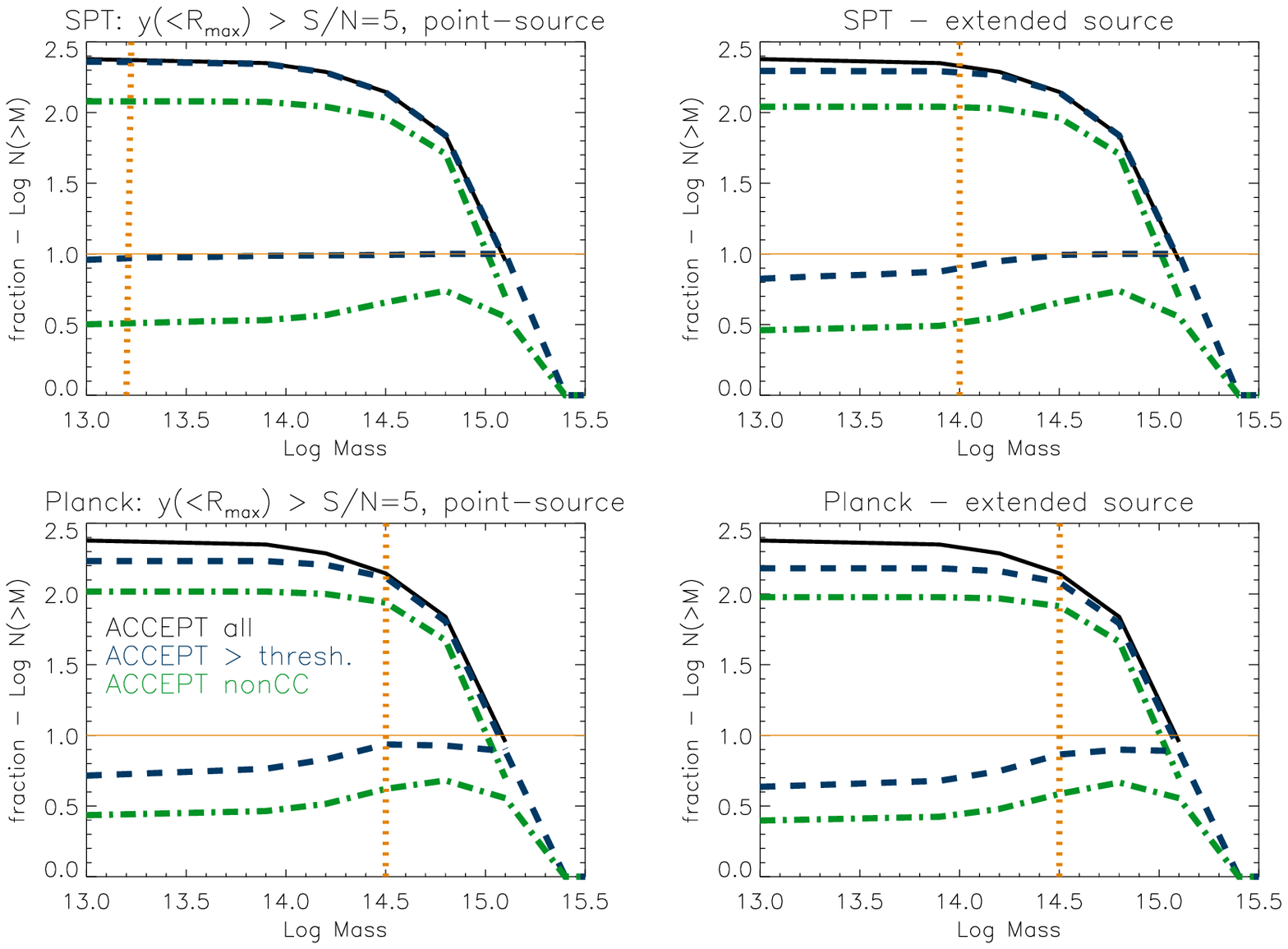}
\caption{Number of clusters with
mass larger than $M$ as a function of M in the ACCEPT sample (solid black
curve), compared to the number of clusters with
mass larger than $M$ whose SZ signal exceeds the detection threshold (dashed line).
We also extract the number of NCC clusters that can be observed (dotted-dashed
line).
In particular we present 2 cases (point-source limit and extended
source) for two different surveys (high resolution - SPT and low
resolution - Planck). In the bottom part of each panel - below the horizontal light solid
curve - we show the ratio of
detected cluster with mass above $M$ to the cumulative
distribution for the entire ACCEPT sample as a dashed line. The
corresponding fraction of detected NCC clusters with respect the total number
of ACCEPT clusters is shown as a dashed-dotted line.
The vertical dotted line marks the mass below which the detection loses
more than the 10\% of objects with respect to the total sample.}
\label{res_accept}
\end{figure*}
\clearpage
\begin{figure}
\includegraphics[width=\textwidth,height=6in]{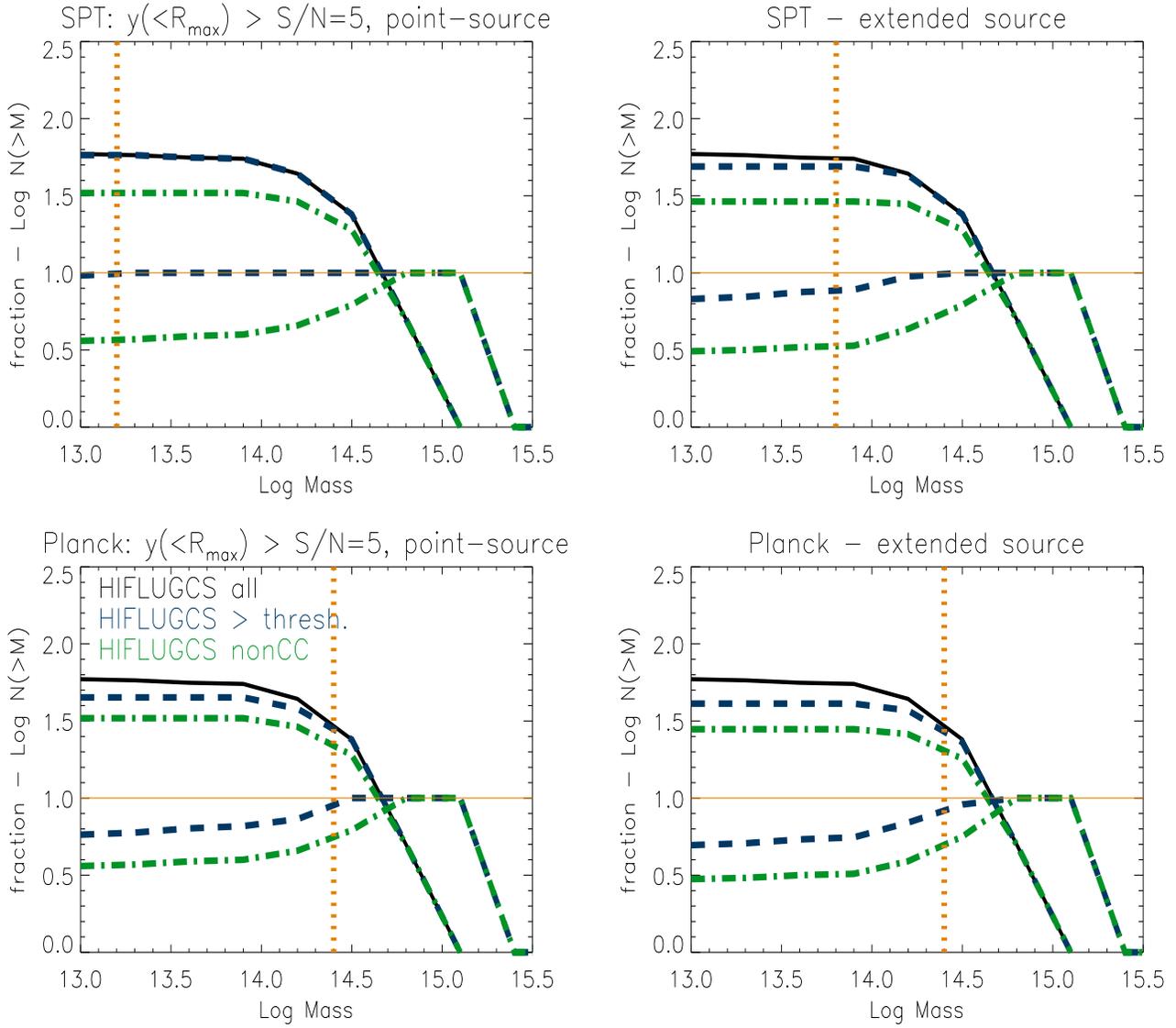}
\caption{As Fig.~\ref{res_accept}, but for the HIFLUGCS subsample.}
\label{res_hiflugcs}
\end{figure}
\clearpage
Here we mention that, in this latter case,} $\theta_c <<$ instrument resolution and the cluster 
detection threshold decreases and tends to the point source case\footnote{Defined
in the Melin et al. (2006) case as the y-axis intercept
of their selection functions, c.f. their Fig.2 .}. According to Melin et al.,
such a detection limit is lower than the extended source case because
of the lower noise associated to the smaller area covered by the cluster.
Therefore, we will also present cases for the point-source detection limit
with the aim of illustrating how the variation in the apparent dimension
of the core changes the detectability of the clusters\footnote{Note that we do not change the cluster redshifts here}.

\section{Results}
\label{res}
\subsection{Overall results}

We first compute how many clusters of our sample can be detected
(given their mass, SZ signal and angular extension) by SPT and Planck.
We wish to stress that an exact determination of the limit mass for cluster detection
by different instrument \emph{is not} the main scope of the paper, because
of the adopted cluster sample characteristics and the fact that we
make use of just one detection method (with its selection function) that
may not be optimal. The values that we will discuss below, instead,
should be taken as the mass at which we focus our study on the CC bias.

In Fig.~\ref{res_accept} we present the number of clusters with
mass larger than $M$ as a function of $M$ in the ACCEPT sample (solid black
curve) and compare it with a similar curve made for all clusters
whose SZ signal exceeds the detection threshold (dashed line).
We also extract the number of NCC clusters that can be observed (dotted-dashed
line).
In particular we present 2 cases (point-source limit and extended
source) for the two different surveys (high resolution - SPT and low
resolution - Planck).
In the bottom part of each panel - below the horizontal light solid
curve - we show the ratio of
detected cluster with mass above $M$ to the cumulative
distribution for the entire ACCEPT sample as a dashed line. The
corresponding fraction of NCC clusters is shown as a dashed-dotted line.
The vertical dotted line marks the mass below which the detection loses
more than the 10\% of \emph{cumulative distribution} of the objects.
Note however, that we will define the limit mass for the detection
when the fraction of clusters \emph{in a given mass bin} falls below
90\%.

Similar plots can be obtained if we perform the integration of the pressure profile out to $R_{500}$
or when for integration of the $\beta$-model profile over 10 $\theta_c$ (as in Melin et al., 2006).
In general, in the point-source approximation SPT recovers all the clusters
above $\sim 2\times 10^{13} M_{\odot}$, whereas when the dimensions of the clusters are
considered the reconstruction can be considered more than 90\% \emph{complete} 
over $10^{14} M_{\odot}$. Again, we stress that an exact determination of completeness
levels of different surveys is beyond the scope of the paper. However, our findings are in line with expectations
from a more thorough assessment of the instrument performances (e.g. Bartlett, 2006)
that took properly into account the number of cluster expected and recovered in a given
volume of the universe.
The point-source limit emphasizes how the finite and non negligible dimensions
of the clusters impact the detection threshold.

In the case of Planck, instead, the performances are such that it
will recover most of the clusters above $\sim 3\times10^{14} M_{\odot}$.
This is not unexpected, since Planck is likely to miss
a certain fraction of relatively massive clusters due to its resolution, as shown by, e.g., Leach et al (2009);
In particular, we refer to this paper - and references therein - and Melin et al. (2006) for a more
accurate description of the completeness of a cluster survey based on Planck's results. 
We stress, however, that even if our sample is not complete and it is not
representative of the volume of the universe that can be probed by a survey like
Planck, the results that we get in our simple approach are consistent
with more detailed calculations.
Furthermore, we double-checked that the ACCEPT mass
distribution (Fig.~\ref{accept_hist}), that peaks at $\sim 5\times10^{14} M_{\odot}$,
does not
introduce some bias around this mass. 

We present the analogous of Fig.~\ref{res_accept} for the HIFLUGCS subsample
(Fig.~\ref{res_hiflugcs}). The performance
estimates given above are substantially unchanged. 
The fraction of recovered NCC clusters with respect to the (either ACCEPT or HIFLUGCS) total sample
does not change, implying that the bias in the recovery of NCC clusters
that we will discuss in the next section is not due to some
feature of the ACCEPT sample nor to the fact that it is not a flux
limited cluster sample.
From the HIFLUGCS subsample we infer that the minimum mass
of the clusters that can be detected (extended source
case) for SPT is $\sim 0.8\times10^{14} M_{\odot}$, the minimum X-ray luminosity
is $0.3\times 10^{44} \rm erg\, s^{-1}$ and the minimum temperature 2.1 keV.
If we limit ourselves to the CC clusters the minimum X-ray luminosity
is somewhat higher, being $0.6\times 10^{44} \rm erg\, s^{-1}$ and possibly corresponding
to the fact the CC clusters are on average more luminous than NCC
clusters at a given temperature.
As far as Planck is concerned, instead, the minimum mass is $\sim 10^{14} M_{\odot}$, the minimum X-ray luminosity is $0.7\times 10^{44} \rm erg\, s^{-1}$ and the minimum temperature 2.8 keV.
Again, if we limit ourselves to the CC clusters, the minimum X-ray luminosity
is somewhat higher, being $0.8\times 10^{44} \rm erg\, s^{-1}$.

{ We wish to remind the reader that the ACCEPT catalog is neither flux limited, nor volume limited. 
According to Cavagnolo et al. (2009), since ACCEPT clusters come from a large number
of observing programs, it is quite unlikely that it is biased toward a particular
class of clusters. However, CC clusters are preferentially high surface brightness
systems, so we cannot exclude that they are over-represented even when we limit
us to the HIFLUGCS subsample.
For instance, optical selection in groups (Rasmussen et al., 2006) hints toward a different
picture than the X-ray selection, in that the fraction of high X-ray surface brightness
systems diminishes. However, groups are below the detection limit of SZ surveys. 
If this were the case also for clusters (Popesso et al., 2007), instead, we should expect that the effect of CCs in SZ cluster
surveys is smaller than found here. A closer look at the X-ray underluminous optically selected 
clusters (Popesso, et al. 2007, Dietrich et al., 2009)
shows that either the mass derived from optical measurements has been over-estimated, or suggests
that these clusters may be still accreting mass and gas from filaments (Bower et al., 1997). In this latter case,
the density and the temperature of the ICM are not high enough to lead to a detection
in X-ray. Such a lack of detectability will be presumably similar (or even stronger, given
the linear dependence of the density) in a SZ survey. While lensing surveys (e.g. Richard et al., 2009) may shed light
on the mass distribution within the cluster and provide mass-limited catalogs, the combination of UV/optical with X-ray/SZ surveys will
certainly allow us to make a more quantitative assessment of the incidence
of CC/relaxed systems (see also Sec. 4.2.3). }

\subsection{Search for a possible bias due to the presence of CC}

From the results of the previous section, it is clear that each survey
is capable of detecting almost all the clusters above a given limit.
We now turn our attention to what happens near this limit and, in particular,
if there is some bias induced by the presence of CCs.
The effect that we expect is multi-fold: i) the CC increases the SZ signal (see Introduction);
ii) CC clusters tend to have smaller core radii, therefore they 
are also favored if the cluster detection technique has a selection
function whose threshold increases with the extension of the cluster core radius
(measured in terms of multiples of the core radius as in Melin et al. 2006); 
iii) if there is not an a-priori knowledge of the cluster mass, this will
be derived by the SZ. Therefore CC clusters with a mass below the nominal
mass of the instrument can make it to be observed 
and they will be assigned a higher mass if their CC status
is not recognized.

\begin{figure}
\includegraphics[width=3.5in,height=3.2in]{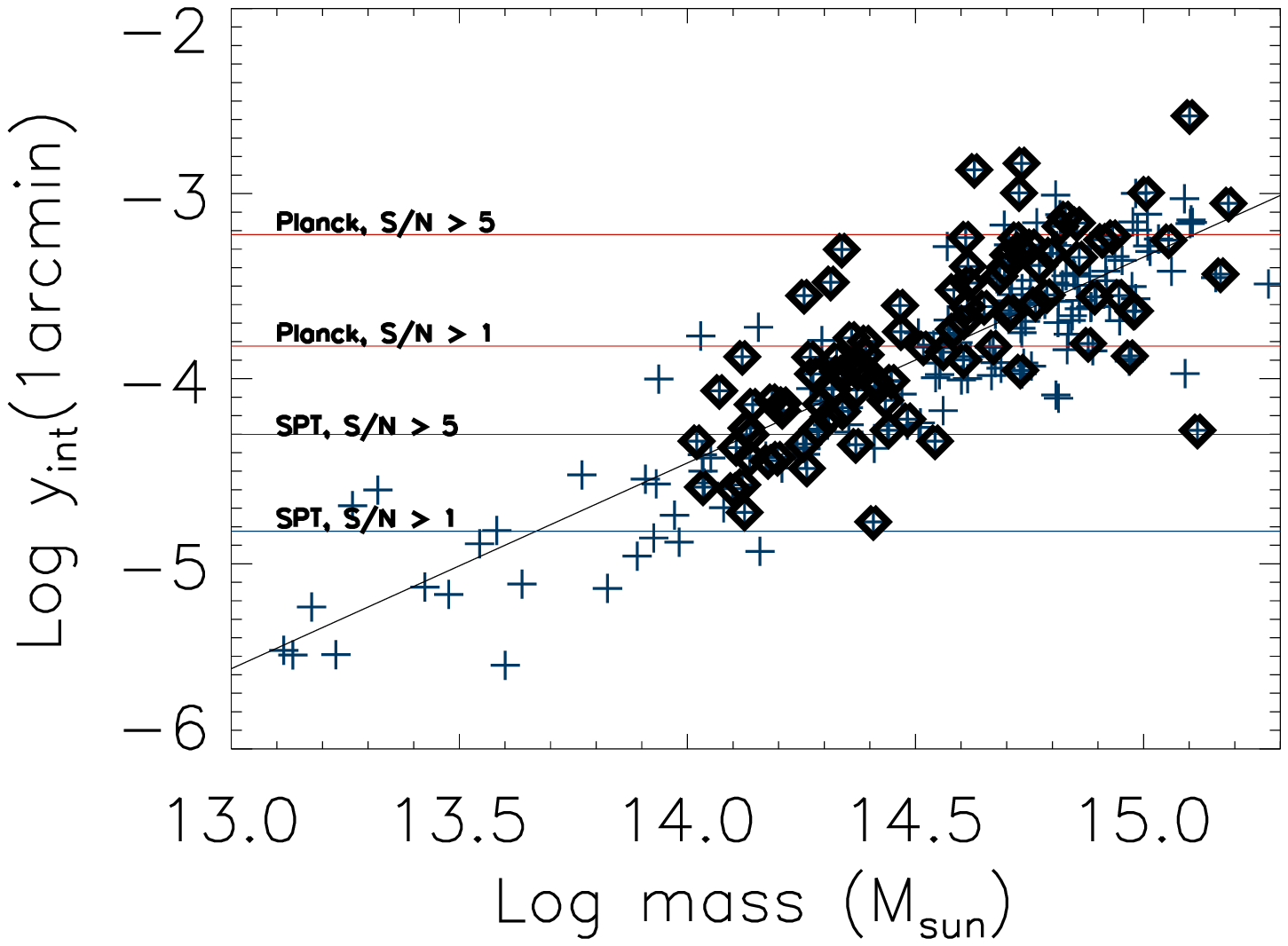}
\includegraphics[width=3.5in,height=3.2in]{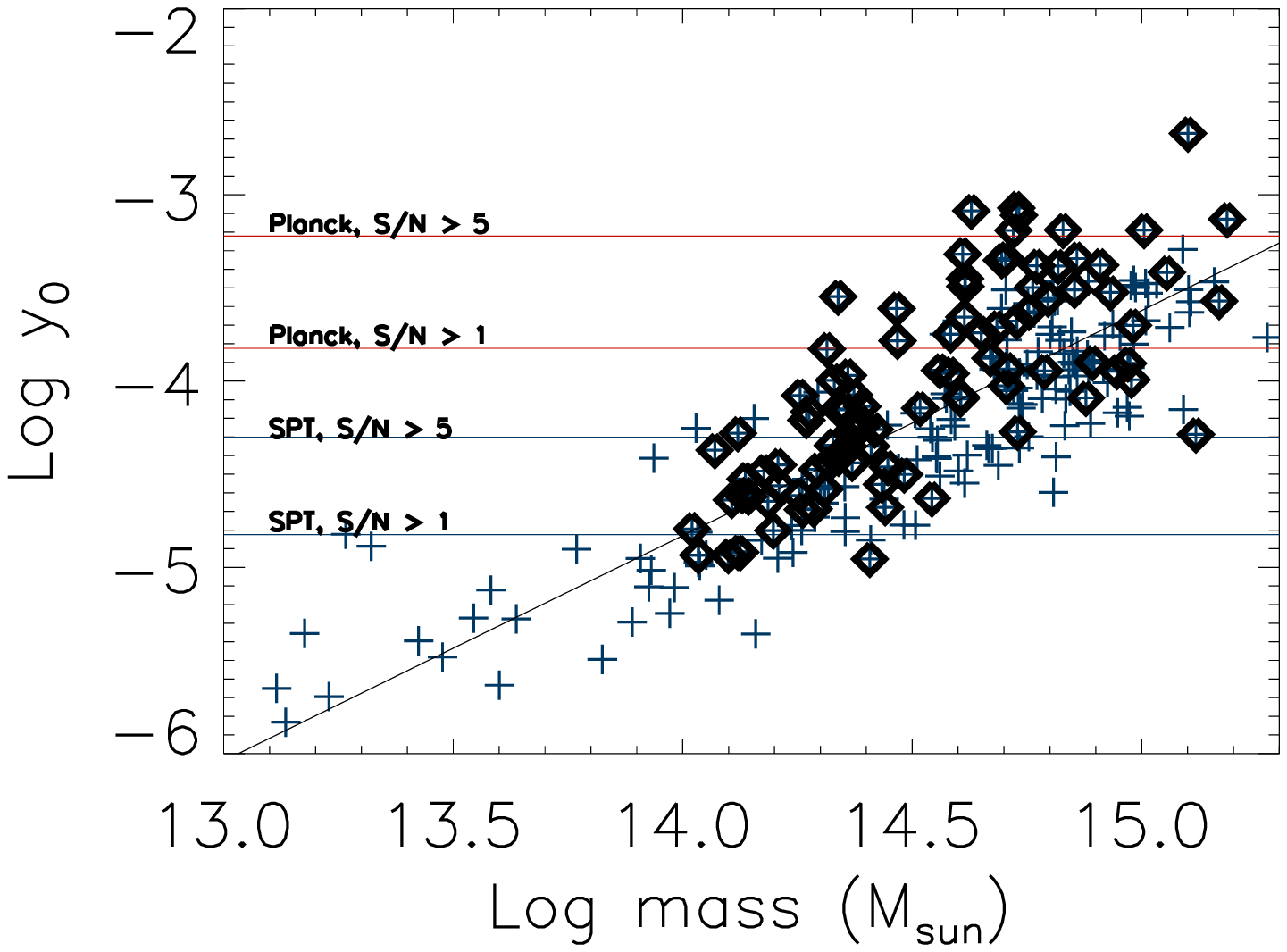}
\caption{Relation between
$y_0$ ($y_{int}(<1arcmin)$ and cluster mass in the entire
ACCEPT sample. Diamonds correspond to CC clusters.
The horizontal lines given the S/N=1 and S/N=5 detection limits
for both SPT and Planck. For the sake of simplicity we use
the point-source detection threshold in order
to emphasize only the effect that the CC has on the predicted SZ signal.}
\label{ymass}
\end{figure}

We limit our investigation on the first one.
In Fig.~\ref{ymass} we show the relation between
$y_0$ (or $y_{int}(<1arcmin)$) and the cluster mass in the entire
ACCEPT sample. The horizontal lines give the S/N=1 and S/N=5 detection limits
for both SPT and Planck. Diamonds correspond to CC clusters
above the lowest of these thresholds.
For the sake of simplicity we use
the point-source detection threshold in order
to emphasize only the effect that the CC has on the predicted SZ signal.
From the bottom panel we infer a clear effect of the CC on $y_0$:
at a given mass, the clusters that display the highest $y_0$ are CC.
We thus validate the suggestions by McCarthy et al. (2003a) on a much
larger sample of clusters than in the McCarthy et al. (2003b) paper.
Were the detection based on $y_0$ we should expect a non-negligible
bias due to the presence of CCs. 
On the other hand, the effect begins to be washed out already at 1arcmin.
Moreover the scatter increases, simply because the clusters are at
different redshifts, therefore the fixed aperture of 1 arcmin
probes differing physical distances.

\begin{figure}
\begin{center}
\includegraphics[width=3in,height=4in]{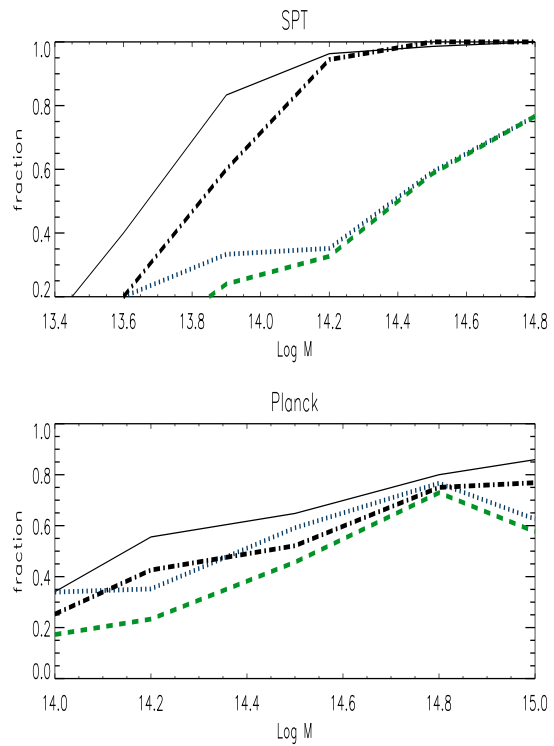}
\caption{Fraction of clusters in
a given mass bin that reached the S/N=5 threshold for 
extended sources in both SPT (upper panel) and Planck (lower
panel) survey (solid lines) in the case in which the signal comes
from the integration of the beta-model out to $R_{500}$. 
 Dot-dashed lines refer to the case in which we use
the SZ signal from the $beta$-model integration for all CC
clusters. The dotted line refers to the \emph{true} fraction of NCC in that mass
bin, whereas the dashed line gives the \emph{observed} fraction of NCC (see text).}
\label{frac_beta}
\end{center}
\end{figure}

In Fig.~\ref{frac_beta} we plot the fraction of clusters in
a given mass bin whose $y_{int}(R_{500})$ reached the S/N=5 threshold for 
extended sources in both SPT (upper panel) and Planck (lower
panel) survey as solid lines. 
Dot-dashed lines refer to the case in which we use
the SZ signal from the $beta$-model integration for all CC
clusters in an effort to mimic the case in which
all clusters were NCC.
We focus on the mass range at which the rate of detections
from the two surveys under study begins to decline. 
One can clearly see that the exclusion of CC regions from clusters
will lower the number of detected clusters. The effect
is around the 20\%, for the SPT case getting larger and larger at lower masses, whereas
it is lower and  more or less constant with for Planck, probably
because of the lower resolution of the latter instrument. 

A more precise estimate of the CC bias that does not rely on a fitted $\beta$ model
can be derived by counting the number of detected NCC clusters and comparing
it to the expected one.
In each panel, the dotted line refers to the ratio of NCC in a given
mass bin to the number of clusters in ACCEPT in the same bin, namely
the \emph{true} NCC fraction\footnote{The \emph{true} CC fraction is simply 1-the \emph{true} NCC fraction.}.
The dashed line, instead, gives the same ratio, but for the NCC that exceed the sensitivity
threshold with respect to the number of ACCEPT clusters \emph{observed}
in that mass bin: this is the \emph{observed} NCC fraction.
We note that the \emph{observed} fraction tracks the true one at high masses,
for both experiments, and then is reduced when the overall
performances of the instruments decrease and 
starts to be significant.
We obtain a bias of 5-15 \%, with a mass evolution that tracks the one
given the dot-dashed Vs. solid curves.
This confirm what we saw in Fig.~\ref{ymass} in the case of the point-source
limit: at a given mass (near the \emph{limit mass}) CC clusters are more likely to be detected.
A sample including more massive
clusters than ours is probably needed to carefully assess the bias in this
Planck case.

{ As noted in the previous section, the X-ray selection may yield
artificially higher fractions of CC clusters.
The fraction of CC clusters in ACCEPT is 40\%, somewhat lower
than the typical $\sim 50-60\%$ reported in the literature (Peres et al., 1998, Chen et al., 2007), 
and decreases at masses above $4\times10^{14} M_{\odot}$ down to the 26\%, in agreement
with both numerical simulations (e.g. Burns et al., 2008) and observations (e.g. Chen et al., 2007).
Therefore our results should be robust as far as the Planck case is concerned. At lower
masses the fraction of CC clusters is more uncertain. Therefore,
for the SPT case, we make the exercise of randomly replacing nearly half of CC clusters
in the mass range $0.5-1.5\times10^{14} M_{\odot}$ with NCC, bringing the fraction of CC clusters in this mass
range down to the 33\%. In particular, the randomly chosen CC clusters have been
assigned 
a SZ signal calculated only from the $\beta$-model
profile. We find that the bias is still present and has a mass dependence similar to what shown in Fig. 8 (upper panel).
The amount is accordingly smaller, however, being around 5\% at $\sim 1\times10^{14} M_{\odot}$.}

\begin{figure}
\begin{center}
\includegraphics[width=3in,height=4in]{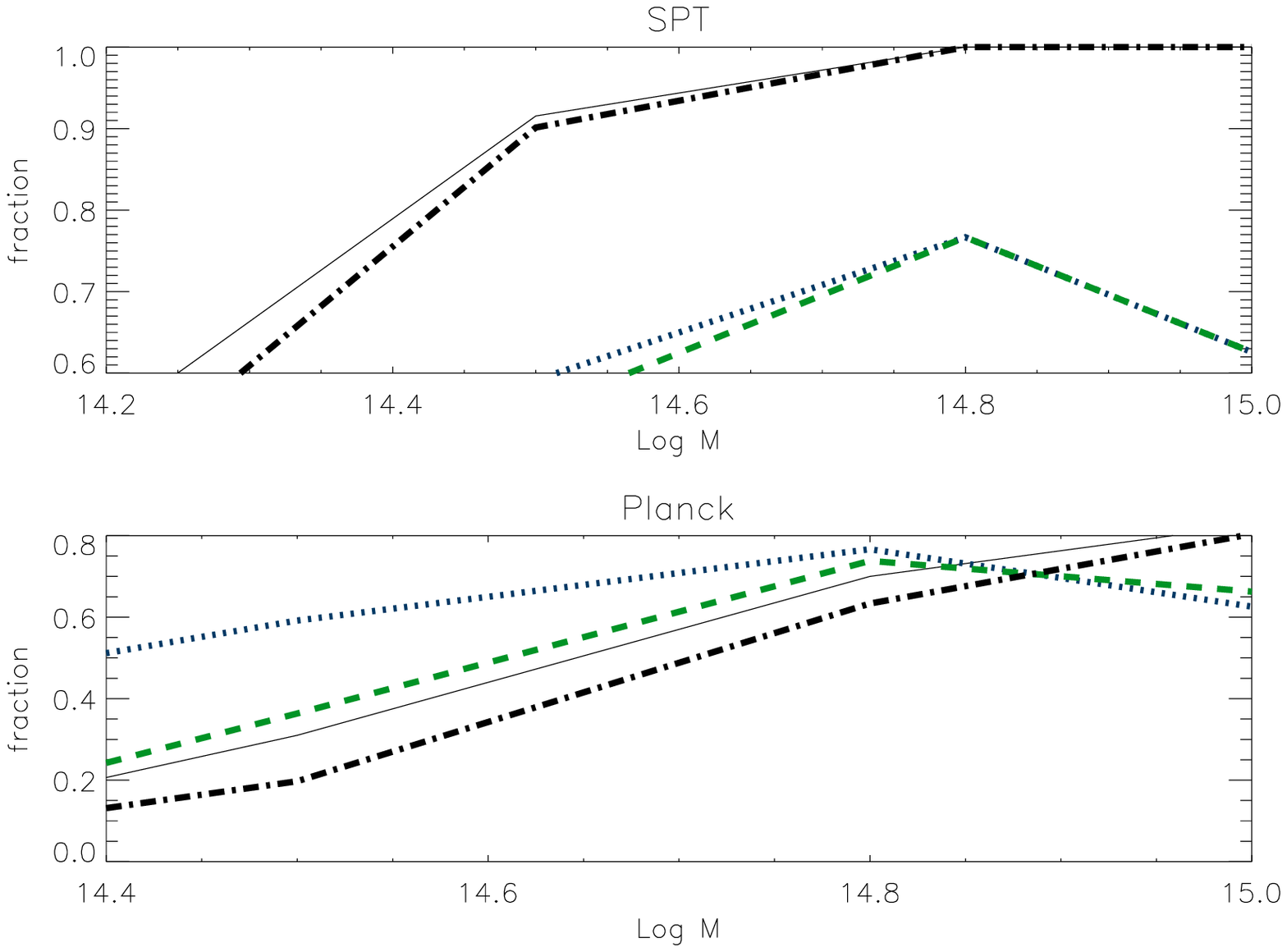}
\includegraphics[width=3in,height=2in]{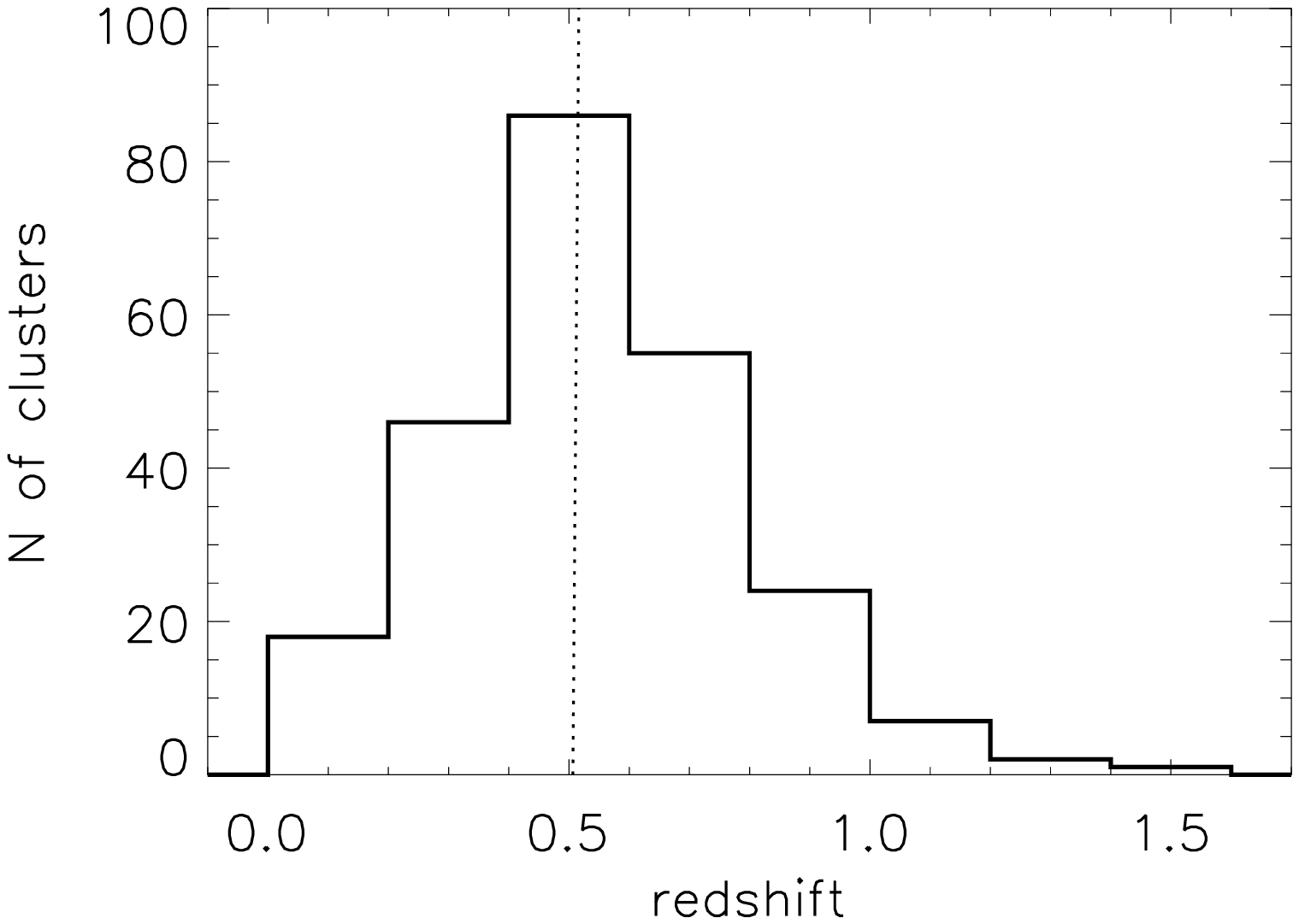}
\caption{Upper panel: fraction of clusters in
a given mass bin that reached the S/N=5 threshold for 
extended sources in both SPT (upper panel) and Planck (lower
panel) survey (solid lines) in the case in which all
clusters are assumed to be at z$\sim$0 (lower panel, see text). Symbols as in Fig.~\ref{frac_beta}.}
\label{frac05}
\end{center}
\end{figure}

{ Another potential concern is that our sample of clusters covers a range in redshift that is not
representative of the expected massive cluster redshift distribution for the current cosmology  (e.g. Halmann et al., 2007).
We present in Fig.~\ref{frac05} (upper panel) the equivalent of Fig.~\ref{frac_beta} for
the case in which all clusters are artificially set at z$\sim$0.5. In particular,
the clusters more massive than $\sim 10^{14} M_{\odot}$ are distributed according to
$dN/dz$ calculated by Halmann et al. (2007) for the WMAP 3 years cosmological parameters.
We show the new redshift distribution in the lower panel of Fig.~\ref{frac05}.
We did not change either the physical dimensions or the mass of these clusters. However,
being the SZ effect integrated on the angle, if all the ACCEPT clusters were at much higher
redshift than they really are their integrated signal over the solid angle will be lower.
Therefore, a sharp decrease in the performances of the surveys is expected and the
minimum recovered mass is higher than in the fiducial case.
Moreover, their $\theta_c$ are accordingly smaller. Since the Melin et al. detection
threshold depends on $\theta_c$, the net effect is lowering the minimum signal needed
to detect them, in that - broadly speaking - more distant cluster can be better approximated as point-sources (
e.g. Fig.5, see also Chamballu et al., 2008). In the SPT case, where the dependence on $\theta_c$ is stronger,
a small amount of bias due to the presence of CC is there, though, at the 5\% level at masses
$\sim 2\times10^{14} M_{\odot}$. For the Planck case, instead,
the results are more similar to what found in  Fig.~\ref{frac_beta}, because the selection function is 
almost constant with $\theta_c$. 

Moreover, while the peak of the distribution of $\sim 10^{14} M_{\odot}$ haloes is around z$\sim$0.5-0.7 for
quite standard cosmological parameters (e.g. Halmann et al., 2007), the peak in the distribution
of detected clusters will be at lower redshifts.
For instance, using the same selection function that we employ (Melin et al),
the expected cumulative number of cluster in the Planck Catalogue
will feature half of the detections at z $<$ 0.3 (e.g. Chamballu et al., 2008, c.f. their
Fig 2), with peak at z$\sim$0.1 (Bartlett et al., 2008).
Therefore the bias is expected to vary as a function of redshift with
the maximum amount being determined by our fiducial case, and the minimum
set by the exercise of shifting all the clusters at z$\sim$0.5.
The exact value can then be assessed by means of a proper simulation of the survey which convolves the expected number of clusters
as a function of the cosmology with the redshift variation of the selection function and of the cluster angular size and cooling
status.}

\section{Discussion}
\label{disc}
In this section we first discuss some potential implications of the bias in SZ-only cluster
detection induced by CC. We also suggest possible diagnostic tools
to understand and pin-point CC clusters in SZ surveys by means
of a multiwavelenght approach. Finally we discuss the implicit assumption
of a non-evolution of the CC cluster fraction with time that is behind
the results presented in the previous section and argue that 
a multiwavelenght approach can have a more general use, for instance to study the evolution of the CC fraction.

\subsection{The bias in the mass estimate}

Several cosmological applications in which SZ-only cluster surveys
are employed need an accurate assessment of the cluster mass - SZ signal
relation. Either this relation is calculated from locally calibrated relations (e.g. Bonamente et al., 2008, 
Morandi et al., 2007) or it is derived from the self-calibration approach (e.g. Majumdar \& Mohr, 2003),
it is important to understand whether the presence of CCs could bias such an estimate.
From the previous sections we understood that, especially around the limit
mass of the instrument, CC clusters are more likely to be detected
than non CC clusters. 
As an example, a linear fit to our calculated SZ signal for clusters at z$<$0.1 and above $2\times10^{14}M_{\odot}$ 
would return $Log (y_{int}(<R_{500})) = 1.86 \cdot Log M_{500} -29.4$ with a standard dispersion of 0.26,
whereas if use limit the analysis to CC clusters we get
$Log (y_{int}(<R_{500})) = 2.02 \cdot Log M_{500} -31.9$ with a standard dispersion of 0.5,
namely the presence of CC steepens the relation, lowers the normalization and induces a larger scatter. 
Observations (e.g. Morandi et al., 2007) have emphasized
that a lower normalization is needed when the cluster sample comprises only CC clusters.
We stress that this is due to the steeper relation obtained for CC clusters only.
Our findings also hold at redshifts above 0.1.

A similar behavior can be shown to hold when one considers the $y_0$-mass relation 
and the $y_{int}(<1arcmin)$-mass relation (see Fig.~\ref{ymass}).
The above example means that it is likely to have CC clusters with a SZ signal a factor of $\sim$2 higher
than the mean value at their given \emph{true} mass. If we tried to assign to such a cluster
an \emph{observational} mass by using the $Log (y_{int}(<R_{max})) -  Log M_{500}$ relation\footnote{
In particular, we use the inverse relation $Log M_{500} = f(y_{int}(<R_{max}))$}
derived for the entire sample, we would obtain that $M_{obs}/M_{true}\sim$ 1.2, namely we
overestimate the mass by $\sim$20\%.
Therefore the presence of CC may induce a ``leak'' of clusters with true masses
around the limiting mass of the instrument towards higher masses
and, at the same time, an inclusion in the observational sample
of systems with a mass slightly lower than the cutoff mass
that without CC would not be observed.

We recall the reader that a more thorough exploration of the consequences
of the CC bias in the recovered mass function would require
a starting sample that follows either a theoretical 
or an observational cluster mass function. As it can
be seen from Fig.~\ref{accept_hist} this is not the case
for our sample.

\subsection{How to recognize the presence of the CC?}

\subsubsection{The case of SZ observations only}

\begin{figure}
\begin{center}
\includegraphics[width=3in]{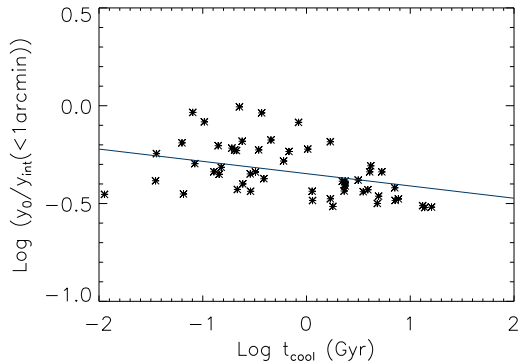}
\caption{Ratio between $y_0$ and $y_{int}(<1arcmin)/arcmin^2$ versus central cooling time
for clusters whose mass is in the range $2-4\times10^{14}M_{\odot}$.}
\label{diag}
\end{center}
\end{figure}

In the previous section we argued that the presence of CC clusters
can bias the SZ-based cluster detection at masses near
the limit mass of the instrument.
It is, therefore, important to understand how to estimate
if such a bias is present in blind SZ surveys.
Interestingly, as shown by Fig.~\ref{diag} for clusters in the mass range $2-4\times10^{14}M_{\odot}$ (i.e.
at nearly the limiting mass for a 90\% complete reconstruction of the cluster
population), the ratio between $y_0$ and $y_{int}(<1arcmin)$
anti-correlates with the central cooling time. In particular,
for NCC clusters (i.e. those with central cooling time above 1 Gyr), this ratio
is always below -0.3, whereas for most of CC clusters is larger than -0.3. 
The plot of Fig.~\ref{diag} can thus be used as a rough diagnostic for
the presence of CCs when only SZ observations are available. In particular, a value of
$Log (y_0/y_{int}(<1arcmin))$ above -0.3 should warn the observer that that particular
cluster might feature a CC and call for follow-up observations.
A similar relation hold when considering the ratio between $y_0$ and $y_{int}(<R_{500})$, albeit
with a larger scatter. Other observational works (e.g. Morandi et al., 2007) have noticed
that the ratio $y_0/y_{int}$ tend to be higher in CC clusters but did not show
any direct evidence of the link with the ICM thermal status.
$y_0$, however, is not directly observed, because current SZ instruments have a much
lower resolution. Therefore, we expect
the CC effect to be blurred in the \emph{observed} ratio $y_0/y_{int}(<1arcmin)$.
In fact,  $y_0$ can only be reconstructed from the observed
signal, often assuming a beta-model, and the result 
might strongly depend on the assumed intracluster gas profile (e.g. Benson et al., 2004).
Interestingly, our finding still holds when considering ratios at different
radii. For instance, the presence of the CC is still evident at half the FWHM of current high-resolution
instruments like SPT, i.e. at 0.5 arcmin, even if $y_0$ cannot be directly observed.
Therefore, we suggest a ratio like $y_{int}(0.5 arcmin)/y_{int}(R_{max})$
as a proxy for the $y_0/y_{int}$ diagnostic of Fig.~\ref{diag}. In this sense, we require next generation instruments to have a minimum resolution of 0.5 arcmin in
order to rely on SZ-only based surveys to understand whether their derived mass
function have a bias induced by CCs.

\subsubsection{X-ray follow-up}

Dubious cases can then be assessed if one has either follow-up or archival X-ray observations.
The cases of in which spectra are available - where a direct estimate of the cooling time is possible - will probably be just a handful.
With X-ray imaging data only, the presence of the CC at high redshift can be the confirmed
by using the ``cuspiness'' of the gas density model used
fitted to the X-ray surface brightness (Vikhlinin, et al., 2006, for a different definition see Santos et al., 2008).
In particular, Santos et al. (2008) derive a power law relation
between the cooling time and an empirical surface brightness concentration index valid out to redshift $\sim 0.9$,
which can be easily combined with the results of our Fig.~\ref{diag}. They do not find a correlation
between the concentration index and the presence of a dominant galaxy. As
we will see in the next section, there is indeed a relation, but
observations at other wavelengths are needed to make it applicable
to the investigation of the CC presence.

\subsubsection{UV and optical follow-up}

For SZ cluster surveys involving a large number of clusters, instead, a cross-correlation with
optically selected clusters can be more beneficial and more effective
than the X-ray follow-up.
Several recent studies have reported
examples of ongoing star formation in the BCGs (see Introduction).
In particular, it seems that up to the 25\% of the BCGs had optical
blue cores (Bildfell et al., 2008), that in many cases render the entire
galactic light bluer than the observed red-sequence (Bower et al., 1992).

Therefore, if CCs make a cluster more visible in the SZ signal, the converse
is true for the optical search based on the \emph{red-sequence}, likely to miss
blue BCGs and, hence, CC clusters.
We thus suggest that a way to enhance the presence of CC clusters is by cross-matching
SZ-based cluster surveys with optical and UV ones
in order to find clusters with a higher than average SZ signal that host unusually blue BCGs.

As an example, if the portion of the sky surveyed in SZ overlaps
with the region observed by the SDSS, one can make use of existing
optical cluster catalogues (Szabo et al, in prep).
In particular, Szabo et al. catalogue is based on a matched filter technique applied
to the SDSS DR6. It comprises of approximately
74000 cluster of galaxies with richness\footnote{measured as the total luminosity
in the r-band within $R_{200}$ in units of $L^*$} above 20 and has been tested to include also blue BGGs.
As shown by Pipino et al. (in prep), this cluster/BCG catalogue is not affected by the colour bias
and can potentially be coupled to SZ surveys
to confirm the presence of CCs.
In such a catalogue, nearly the 15\% of BCGs in clusters with richness 30 is well below (i.e. bluer than) the red-sequence
at their given redshift, and likely to be missed by red-galaxy search only. 
Remarkably this happens
in the 10\% of the richest clusters (i.e. richness above 50), which will be surely observed in SZ survey.
By using the richness-temperature relation derived by Szabo et al. (in prep) and
the mass-temperature adopted in this paper, a
richness of 50 roughly corresponds
to a mass of $\sim 10^{14} M_{\odot}$, namely where the CC bias seems to be most effective (see previous sections). 
We expect the clusters hosting very blue BCGs to harbor the strongest CC and thus to be 
the ones that deviate more from the $y_{int}$-mass relation.
From the BCG analysis, we derive that the blue BCG fraction remains constant in the redshift
range 0.1-0.4. In the same redshift range the CC fraction seems to stay constant as well
around the value of 50\% (e.g Bauer et al., 2005). Therefore the cross-matching
of SZ and optical surveys can emphasize 1/5 of the CC clusters, presumably
the strongest ones.

\begin{figure}
\begin{center}
\includegraphics[width=3in]{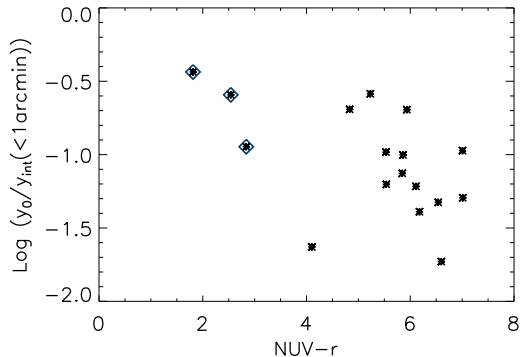}
\caption{Ratio between $y_0$ and $y_{int}/arcmin^2$ versus the NUV-r colour
for BCGs in clusters where combined SDSS and GALEX photometry is available (asterisks) above a mass of $\sim 2\cdot 10^{14} M_{\odot}$. 
Galaxies hosted in CC clusters are emphasized with the diamond symbol. Note that UV-optical \emph{blue} galaxies are those
with NUV-r below 4.}
\label{diag1}
\end{center}
\end{figure}

In reality, the recent star formation creates blue cores which not necessarily affect
the entire optical colour of the galaxy. Nonetheless, it could be enough to make
the galaxy blue in the UV-optical colours. Indeed, Pipino et al. (2009a) showed
a one to one correspondence between blue cores and UV-optical colours for a limited number
of BCGs. Here we assume that this link applies to the entire population of BCGs.
Both ways to infer the presence of a CC hold true in the ACCEPT sample for the clusters
which have a counterpart in the Szabo et al. catalogue (Pipino et al., in prep).
In our specific case of the ACCEPT sample, a cross match between the
Szabo et al. catalogue returns 68 clusters in common. In Pipino
et al. (in prep.) we show that optically blue BCGs are hosted by CC clusters. 
In particular, the NUV-r colour correlates with the excess entropy of these clusters:
the smaller the NUV-r colour, the lower the entropy and the shorter the cooling time.
In this way it will be possible to pick up CC clusters that cannot be 
detected or confirmed by means of the optical colours only.
We refer the reader to Pipino et al. (2009, in prep) for a more detailed discussion on the colours of the BCGs in relation
to the state of the intracluster medium of the host cluster.
Here we show (Fig.~\ref{diag1}) how the NUV-r colours of the BCG can be a proxy
for the ratio $y_0/y_{int}$.
Unfortunately only a limited number of BCGs in clusters with mass above $\sim 2\cdot 10^{14} M_{\odot}$
has both SDSS and Galex imaging. However, we can clearly see that that UV-optical \emph{blue} galaxies, namely those
with NUV-r below 4 (asterisks encircled by a diamond), stand out in a region of relatively high values for the ratio $y_0/y_{int}$,
whereas more \emph{normal} galaxies tend to have a lower $y_0/y_{int}$. BCGs on the NUV-r red sequence (i.e.
those with NUV-r above 6) display the lowest values for $y_0/y_{int}$.
{ If BCGs undergo pure passive evolution at redshifts below z=1, and taking into
account the Galex Medium Imaging Survey (MIS) magnitude limit (Martin et al., 2007), we expect to easily detect UV-blue BCGs out to z$\sim$0.6 with current
instruments. Moreover, if the Galex MIS were continued and extended to have a larger overlap with, e.g., SDSS,
we might be able to: i) confirm the findings of Fig. 11 on the basis of a larger number of BCGs; and ii) 
probe the region where recent observations suggest a strong decline
of the CC fraction (see below).}

\subsubsection{H$_{\alpha}$, IR and radio follow-up}

\begin{figure}
\begin{center}
\includegraphics[width=3in]{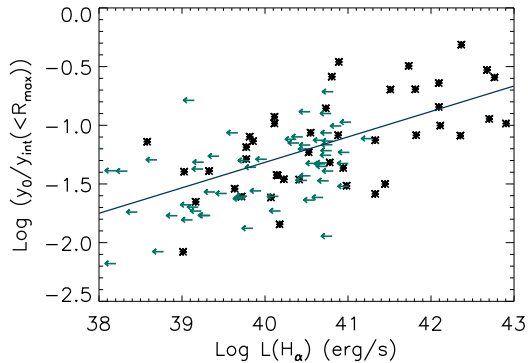}
\caption{Ratio between $y_0$ and $y_{int}/arcmin^2$ versus the H$_{\alpha}$
luminosity as provided in the ACCEPT database. The arrows are upper limit on the luminosity and typically
they represent clusters with entropy above 30 keV cm$^2$. The solid line is a linear regression
fit to the points (excluding the upper limits).}
\label{diag2}
\end{center}
\end{figure}

The ACCEPT catalogue is complemented by H$_{\alpha}$ and radio luminosities associated to the BCGs
in roughly half of the total sample.
As for the former, they are derived from a number of sources (Crawford et al., 1999, Cavagnolo et
al. 2008a, 2009 and references therein) and several apertures that do not
reflect the actual full H$_{\alpha}$ flux from the galaxy. The latter measurements,
instead, come from the NRAO VLA sky survey 
and the Sydney
University Molonglo sky survey ( 
refer to Cavagnolo
et al. 2008a for further details and references).
Indeed, Cavagnolo et al. (2008, see also Donahue et al., 2005) showed that below the entropy threshold
of 30 keV cm$^2$, both the H$_{\alpha}$ and the radio emission suddenly
\emph{switches on}, whereas galaxies in clusters above that limit are quiescent
and only upper limits are detected. In practice, such as emission can be either 
traced back to the recent star formation or to the AGN feedback (Edwards et al., 2009).
Edwards et al. (2007) further
showed that H$_{\alpha}$ emission, while present in 20\% of the entire population BCGs, 
systematically appears in BCGs bluer than expected. 
In this section we make use of the above findings and of the luminosities
provided in the ACCEPT database to suggest other diagnostic tools to
infer the presence of a CC.
In particular, in Fig.~\ref{diag2}, we show the ratio between $y_0$ and $y_{int}$ versus the H$_{\alpha}$
luminosity as provided in the ACCEPT database. The arrows are upper limit on the luminosity and typically
they represent clusters with entropy above 30 keV cm$^2$. The solid line is a linear regression
fit to the points (excluding the upper limits). We find that $Log (y_0/y_{int}(<1arcmin)/arcmin^2 \sim 0.2 log L(H_{\alpha})$. While a proper estimate of the slope of such relation would require a careful assessment
of the aperture effects in the H$_{\alpha}$ measurements, our finding emphasizes the fact
that  H$_{\alpha}$ emission is a clear indicator of low $K_0$ and, hence, a strong CC status. 
Since H$_{\alpha}$ emission is detected in one quarter of the clusters (Crawford et al., 1999), as opposite to the 10\% of clusters hosting optically blue BCGs, such information is helpful to confirm the CC status
in a wider range of cases than the cross matching process with optical surveys.
The drawback of such proxy for the CC is that requires either spectral information or narrow-band imaging,
whereas large surveys as the SDSS adopt larger bands for imaging and do not have spectral information
for all the detected sources.

Interestingly, O'Dea et al. (2008) showed a correlation between IR emission measured by the Spitzer mission  and literature H$_{\alpha}$ emission in BCGs, both
being caused by star formation in most of the cases. Since O'Dea et al. (2008, see also Egami et al. 2006)
show an anti correlation between IR luminosities and cooling time, we do not repeat the exercise
here and take their findings as corroborative to our argument.

\begin{figure}
\begin{center}
\includegraphics[width=3in]{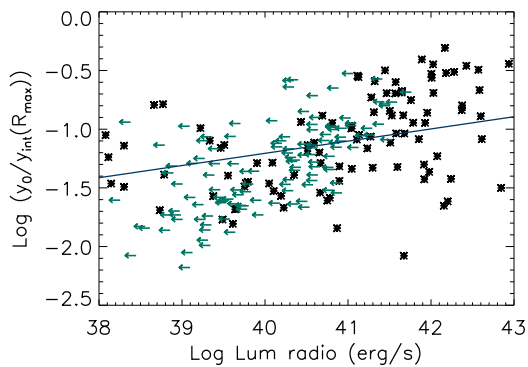}
\caption{Ratio between $y_0$ and $y_{int}/arcmin^2$ versus the radio
luminosity as provided in the ACCEPT database.The arrows are upper limit on the luminosity and typically
they represent clusters with entropy above 30 keV cm$^2$. The solid line is a linear regression
fit to the points (excluding the upper limits).}
\label{diag3}
\end{center}
\end{figure}

Similar considerations apply to the radio luminosities (Fig.~\ref{diag3}, see also Mittal et al., 2009). In this latter case the relation
is flatter, being $Log (y_0/y_{int}) \sim 0.1 log L(radio)$, and exhibits a larger 
scatter than in the former. Incidentally, we note that radio sources associated with BCGs seem to have a steeper
spectral index than sources associated with other cluster members (Lin \& Mohr, 2007), thus making the detection of
the former more robust.
However, in the radio case, the fact that the luminosity is presumably associated to the AGN - rather than
to recent star formation - implies
that the AGN duty cycle cannot be neglected. Moreover, a number of clusters with $K_0$ larger than 30 keV cm$^2$
and an active AGN in the BCG are present in the ACCEPT sample.
In a sense, a diagnostic based on the radio
is less informative on the recent star formation and the CC status.
Another drawback of the adopted radio surveys is that their current resolution is such 
that at redshift larger than 0.2 it is difficult to assess whether the radio
emission is actually linked to the BCG.
However, a multiwavelenght approach involving UV, optical, IR and radio bands
can put firm constraints on the CC status and make emission lines
and unusual colours in BCGs unambiguously linked to either recent star formation episodes
or AGN activity.

\subsection{The effect of the evolution of CC and BCG properties at z$>$0.5}
\label{evo}

{ An important caveat that the reader should bear in mind is that,
until this point, we have implicitly assumed that the CC fraction does
not evolve with time. Indeed, Bauer et al. (2005) presented a comparative study
of cool cores at low and intermediate redshift and found 
no signs of evolution of the cool core fraction in the redshift range 0.15 - 0.4 to the present time.
This corroborates our assumption and makes the method for detecting CC clusters
via UV/optical colours presented in the previous sections robust
in the redshift range 0.1-0.4.
At higher redshifts, where high resolution SZ surveys are expected to
deliver a fraction of clusters, it seems that { strong} CCs disappear 
(e.g. Vikhlinin et al., 2006, Santos et al. 2008), therefore the CC-induced bias might quickly vanish accordingly. 
In particular, while Vikhlinin et al. suggest that the CC disappear at z$>0.5$, the analysis by Santos et al., 
indicates a significant fraction of distant clusters harboring a moderate CC out to z=1.4, similar to the local fraction,
and a decrease in strong CC at z$>0.7$. Both works argue that 
the absence of strong cooling is likely linked to the higher merger rate expected at these redshifts, 
and should also be related to the shorter age of distant clusters, implying less time to develop a cool core.
The role of mergers in destroying cool cores has long been
debated (Fabian et al. 1984), 
both from observational results and from simulations. Currently, observations seem to favor
CC destruction through cluster mergers (Allen et al. 2001,
Sanderson et al. 2006). Simulations, however, yield ambivalent
results:
CC can be destroyed by early major mergers according to Burns et al. (2008), whereas
Poole et al. (2006) argue that, in the CDM scenario, the
merger rate is too small to account for the local abundance of
non-CC clusters whose disruptions occurs on short timescales. { Interestingly, the simulations by Burns et al. (2008) do not
show the decrease of CC clusters at z$>$0.7.} Therefore, a quantitative study
of the CC evolution in the redshift range 0.5--0.7 is crucial for these
studies as well as for understanding the properties of SZ-only selected
clusters. In a sense, we can turn the argument around and suggest that the CC evolution might be independently
confirmed } by cross-matching SZ-based cluster catalogues with UV/optically/IR selected ones,
especially at the high richness end, namely clusters that are detectable
out to the highest redshifts.
In the previous section we have seen that the fraction of rich CC clusters hosting a blue BCG
is around 10\% in the redshift range 0.1-0.4. Observational studies (e.g. Stanford et al., 1998)
suggest that the red-sequence evolution out to redshift 1 is consistent with a pure passively
evolution (i.e. no major star formation episodes) and does not depend from the cluster
richness or its X-ray properties, especially at the high mass end (e.g. Bundy et al., 2005).
{ No evolution in the number density of massive early type galaxies is detected
in the same redshift range (e.g. Scarlata et al., 2007). 
On the theoretical side remarkably either models
based on monolithic collapse (e.g. Pipino \& Matteucci 2004) or those based on the hierarchical growth
of structure (Pipino et al., 2009b, De Lucia \& Blaizot, 2007) predict basically no
major star formation episodes involving BCGs at z$<$ 1 and a very mild evolution of the colours.
If there is any residual star formation, this must be associated with some ``cooling flow'' not
entirely offset by the heating sources. On example might be the cluster XMMU J1229+0151(Santos et al., 2009) at z$\sim$1, 
where some residual star formation is inferred from emission lines in an otherwise normal (optically red) BCG.}
 Therefore, in the simplest scenario in which BCGs do not
undergo mergers at redshift z$<$1, a complete UV/optical + SZ cluster survey would exhibit
a fraction of rich clusters featuring an excess in the SZ signal and a (UV/optical) blue BCG gradually decreasing
from the 10\% to zero if a CC evolution were in place at  z = 0.5--1.
More accurate predictions, however, require
a careful treatment of the colour classification of the BCGs at z$>$0.5, 
with a detailed consideration of all evolutionary effects and the errors in the photometry, 
that are beyond the scope of this paper.
{ Here we just note that the Butcher \& Oemler (1978) effect, i.e. the increase with redshift of the fraction of blue cluster galaxies, 
seems strictly connected to the decrease of the fraction of cluster S0 galaxies with redshift 
(Dressler et al. 1997; Poggianti et al. 1999; van Dokkum et al. 2000; Fasano et al. 2000; Tran et al. 2005),
whereas the fraction of elliptical galaxies - which dominate the central regions of galaxy clusters - does not change. 
Therefore these effect is likely to affect only a minor fraction of the "early type" galaxies as a whole. More important, 
it seems that the trasformation of late-type galaxies into S0 is not restricted to, and possibly even avoids, the cores
(see Wilman et al 2009 in the case of optically selected groups). This seems to be the case of higher redshift clusters,
e.g. MS 2053-04 at z$\sim$0.6 (Tran et al., 2005), in that the properties of the blue cluster 
members in both the main cluster and infalling structures indicate that they will evolve into low-mass 
L$<$L$*$ galaxies with extended star formation histories like that of low-mass S0 galaxies in Coma, making
us confident that the properties of the BCG will not be affected. }

\section{Conclusions}
\label{conc}

We used the high quality pressure profiles of 239 galaxy cluster made
available by the ACCEPT project (Cavagnolo et al. 2009) in order
to derive the expected SZ signal in a variety of cases
that hardly find a counterpart in the simulations.
We made use of the Melin et al. (2006) estimates
for the detection threshold in the cases of point-like
and extended sources for the SPT and the Planck instruments.
Prior knowledge of the entropy profiles of the same clusters
allows us to study the impact of CC in cluster detection
via SZ experiment and to test if this might introduce a bias
in inferred quantities as, e.g., the cluster mass function.

We infer a clear effect of the CC on $y_0$.
We thus validate the suggestions by McCarthy et al. (2003) on a much
larger sample of clusters.
For a high resolution experiment like SPT, we expect that
the fraction of detected clusters with respect to the total
to decline at masses around $\sim 2\times10^{14} M_{\odot}$.
For Planck this happens at a somewhat higher mass.
We found that the presence of CC introduces a small bias in cluster
detection, especially around the mass at which the 
performances of the survey begins to significantly decrease.
If the CC were removed, a lower overall fraction
of detected clusters would be expected.
In order to estimate the presence of such a bias by means
of SZ and X-ray observations, we show that the ratio $y_0/y_{int}$
anti-correlates with the central cooling time.
If multi-band optical cluster surveys are either available for a cross-match
or a follow-up is planned, we suggest that likely CC clusters
are those with a BCG at least 0.3 magnitudes bluer than the average.
Using the Szabo et al. (2009) SDSS DR6 cluster catalogue, we estimate
that this should happen in at least the 15\% of rich clusters.
A more robust estimate of the CC presence is given by
UV-optical colours, like the NUV-r, whose values can be 4 magnitudes
off the NUV-r equivalent of the red sequence, in clusters with low
excess entropy.
We argue that the analysis of a combined SZ/optical/UV survey can be also used
to shed light on the suggested CC evolution with redshift.



\section*{Acknowledgments}
The authors thank the referee for 
his insightful comments that improved the quality of the paper.
AP wishes to thank S.Ameglio for many enlightening discussions, I.Balestra, J.B. Melin and P.Tozzi for
useful comments, K. Cavagnolo for clarifications on the ACCEPT sample,
AP and EP acknowledge support from NSF grant AST-0649899.
EP is also supported by NASA grant NNX07AH59G and JPL-Planck subcontract 1290790
and acknowledges the hospitality of the Aspen Center for physics during
the completion of this work.
This research has made use of: the SIMBAD database, operated at CDS, Strasbourg, France;
the X-Rays Clusters Database (BAX), which is operated by the Laboratoire d'Astrophysique de Tarbes-Toulouse (LATT),
under contract with the Centre National d'Etudes Spatiales (CNES); the NASA/IPAC Extragalactic Database (NED) which is operated by the Jet Propulsion Laboratory, California Institute of Technology, under contract with the National Aeronautics and Space Administration.



\begin{thebibliography}{99}

\bibitem[\protect\citeauthoryear{Allen 
\& Fabian}{1998}]{1998MNRAS.297L..57A} Allen S.~W., Fabian A.~C., 1998, MNRAS, 297, L57 


\bibitem[\protect\citeauthoryear{Allen, Ettori, 
\& Fabian}{2001}]{2001MNRAS.324..877A} Allen S.~W., Ettori S., Fabian A.~C., 2001, MNRAS, 324, 877 




\bibitem[\protect\citeauthoryear{Arnaud, Pointecouteau, 
\& Pratt}{2005}]{2005A&A...441..893A} Arnaud M., Pointecouteau E., Pratt G.~W., 2005, A\&A, 441, 893 


\bibitem[\protect\citeauthoryear{Arnaud, Aghanim, 
\& Neumann}{2002}]{2002A&A...389....1A} Arnaud M., Aghanim N., Neumann D.~M., 2002, A\&A, 389, 1 


\bibitem[\protect\citeauthoryear{Barbosa et 
al.}{1996}]{1996A&A...314...13B} Barbosa D., Bartlett J.~G., Blanchard A., Oukbir J., 1996, A\&A, 314, 13 

\bibitem[\protect\citeauthoryear{Bartlett}{2006}]{2006astro.ph..6241B} 
Bartlett J.~G., 2006, astro, arXiv:astro-ph/0606241 

\bibitem[\protect\citeauthoryear{Bartlett et 
al.}{2008}]{2008AN....329..147B} Bartlett J.~G., Chamballu A., Melin J.-B., 
Arnaud M., Members of the Planck Working Group 5, 2008, AN, 329, 147 



\bibitem[\protect\citeauthoryear{Bauer et al.}{2005}]{2005MNRAS.359.1481B} 
Bauer F.~E., Fabian A.~C., Sanders J.~S., Allen S.~W., Johnstone R.~M., 
2005, MNRAS, 359, 1481 


\bibitem[\protect\citeauthoryear{Benson, Reichardt, 
\& Kamionkowski}{2002}]{2002MNRAS.331...71B} Benson A.~J., Reichardt C., Kamionkowski M., 2002, MNRAS, 331, 71 


\bibitem []{}Bildfell, C., Hoekstra, H., Babul, A., \& Mahdavi, A. 2008, MNRAS, 389, 1637
\bibitem[\protect\citeauthoryear{Bonamente et 
al.}{2008}]{2008ApJ...675..106B} Bonamente M., Joy M., LaRoque S.~J., 
Carlstrom J.~E., Nagai D., Marrone D.~P., 2008, ApJ, 675, 106 


\bibitem[\protect\citeauthoryear{Borgani et 
al.}{2002}]{2002MNRAS.336..409B} Borgani S., Governato F., Wadsley J., 
Menci N., Tozzi P., Quinn T., Stadel J., Lake G., 2002, MNRAS, 336, 409 

\bibitem []{}Bower, R.G., Lucey, J.R., Ellis, R.S., 1992, MNRAS, 254, 589

\bibitem[\protect\citeauthoryear{Bower et al.}{1997}]{1997MNRAS.291..353B} 
Bower R.~G., Castander F.~J., Ellis R.~S., Couch W.~J., Boehringer H., 
1997, MNRAS, 291, 353 

\bibitem []{}Bundy, K., Ellis, R.S., \& Conselice, C.J. 2005, ApJ, 625, 621

\bibitem[\protect\citeauthoryear{Burns et al.}{2008}]{2008ApJ...675.1125B} 
Burns J.~O., Hallman E.~J., Gantner B., Motl P.~M., Norman M.~L., 2008, 
ApJ, 675, 1125 

\bibitem[\protect\citeauthoryear{Butcher 
\& Oemler}{1978}]{1978ApJ...219...18B} Butcher H., Oemler A., Jr., 1978, ApJ, 219, 18 




\bibitem []{} Cardiel, N., Gorgas, J.; Aragon-Salamanca, A., 1998, MNRAS, 299, 977
\bibitem[\protect\citeauthoryear{Carlstrom, Holder, 
\& Reese}{2002}]{2002ARA&A..40..643C} Carlstrom J.~E., Holder G.~P., Reese E.~D., 2002, ARA\&A, 40, 643 

\bibitem []{} Cavagnolo, K.W., Donahue, M.,Voit, G.M.,  Sun, M., 2008, ApJ, 682, 821
\bibitem[\protect\citeauthoryear{Cavagnolo et 
al.}{2009}]{2009ApJS..182...12C} Cavagnolo K.~W., Donahue M., Voit G.~M., 
Sun M., 2009, ApJS, 182, 12 


\bibitem[\protect\citeauthoryear{Cavagnolo et 
al.}{2008}]{2008ApJ...683L.107C} Cavagnolo K.~W., Donahue M., Voit G.~M., 
Sun M., 2008, ApJ, 683, L107 (2008a)
\bibitem[\protect\citeauthoryear{Cavaliere 
\& Fusco-Femiano}{1978}]{1978A&A....70..677C} Cavaliere A., Fusco-Femiano R., 1978, A\&A, 70, 677 

\bibitem[\protect\citeauthoryear{Chamballu et 
al.}{2008}]{2008arXiv0805.4361C} Chamballu A., Bartlett J.~G., Melin J.~-., 
Arnaud M., 2008, arXiv, arXiv:0805.4361

\bibitem[]{}Chen, Y.; Reiprich, T. H.; Bohringer, H.; Ikebe, Y.; Zhang, Y.-Y. 2007 A\&A, 466, 805
\bibitem []{}Crawford C. S., Allen S. W., Ebeling H., Edge A. C., Fabian A. C., 1999, MNRAS, 306, 857
\bibitem []{}De Lucia, G, \& Blaizot, J. 2007, MNRAS, 375, 2
\bibitem[\protect\citeauthoryear{Diego et al.}{2002}]{2002MNRAS.336.1351D} 
Diego J.~M., Vielva P., Mart{\'{\i}}nez-Gonz{\'a}lez E., Silk J., Sanz 
J.~L., 2002, MNRAS, 336, 1351 

\bibitem[]{}Dietrich, J. P.; Biviano, A.; Popesso, P.; Zhang, Y.-Y.; Lombardi, M.; Bšhringer, H.	2009 A\&A, 499, 669





\bibitem[\protect\citeauthoryear{Delabrouille, Melin, 
\& Bartlett}{2002}]{2002ASPC..257...81D} Delabrouille J., Melin J.-B., Bartlett J.~G., 2002, ASPC, 257, 81 

\bibitem[\protect\citeauthoryear{Dietrich et 
al.}{2009}]{2009A&A...499..669D} Dietrich J.~P., Biviano A., Popesso P., Zhang Y.-Y., Lombardi M., B{\"o}hringer H., 2009, A\&A, 499, 669 


\bibitem[\protect\citeauthoryear{Donahue et 
al.}{2005}]{2005ApJ...630L..13D} Donahue M., Voit G.~M., O'Dea C.~P., Baum 
S.~A., Sparks W.~B., 2005, ApJ, 630, L13 

\bibitem[\protect\citeauthoryear{Dressler et 
al.}{1997}]{1997ApJ...490..577D} Dressler A., et al., 1997, ApJ, 490, 577
\bibitem []{}Edge A. C., 2001, MNRAS, 328, 762
\bibitem []{}Edge A. C., Wilman R. J., Johnstone R. M., Crawford C. S., Fabian A. C., Allen S. W., 2002, MNRAS, 337 ,49

\bibitem[\protect\citeauthoryear{Edwards et 
al.}{2009}]{2009MNRAS.396.1953E} Edwards L.~O.~V., Robert C., Moll{\'a} M., 
McGee S.~L., 2009, MNRAS, 396, 1953 


\bibitem[\protect\citeauthoryear{Edwards et 
al.}{2007}]{2007MNRAS.379..100E} Edwards L.~O.~V., Hudson M.~J., Balogh 
M.~L., Smith R.~J., 2007, MNRAS, 379, 100 


\bibitem []{}Egami E., Misselt K. A., Rieke G. H., Wise M. W., Neugebauer G., Kneib J.-P., Le Floc'h E., Smith G. P., Blaylock M., Dole H., Frayer D. T., Huang J.-S., Krause O., Papovich C., Perez-Gonzalez P. G., Rigby J. R., 2006, ApJ, 647, 922
R.M., 1997, ApJ, 483, 582

\bibitem[\protect\citeauthoryear{Fabian et al.}{1994}]{1994MNRAS.267..779F} 
Fabian A.~C., Crawford C.~S., Edge A.~C., Mushotzky R.~F., 1994, MNRAS, 
267, 779 

\bibitem[\protect\citeauthoryear{Fasano et al.}{2000}]{2000ApJ...542..673F} 
Fasano G., Poggianti B.~M., Couch W.~J., Bettoni D., Kj{\ae}rgaard P., 
Moles M., 2000, ApJ, 542, 673

\bibitem []{}Fisher, D., Franx, M., Illingworth, G. 1995, ApJ, 448, 119
\bibitem []{}Goto T., 2005 MNRAS, 360, 322
\bibitem[\protect\citeauthoryear{Hallman et 
al.}{2007}]{2007ApJ...665..911H} Hallman E.~J., Burns J.~O., Motl P.~M., 
Norman M.~L., 2007, ApJ, 665, 91

\bibitem[\protect\citeauthoryear{Herranz et 
al.}{2002}]{2002MNRAS.336.1057H} Herranz D., Sanz J.~L., Hobson M.~P., 
Barreiro R.~B., Diego J.~M., Mart{\'{\i}}nez-Gonz{\'a}lez E., Lasenby 
A.~N., 2002, MNRAS, 336, 1057 


\bibitem[\protect\citeauthoryear{Herranz et 
al.}{2002}]{2002ApJ...580..610H} Herranz D., Sanz J.~L., Barreiro R.~B., 
Mart{\'{\i}}nez-Gonz{\'a}lez E., 2002, ApJ, 580, 610 

\bibitem[\protect\citeauthoryear{Holder et al.}{2000}]{2000ApJ...544..629H} 
Holder G.~P., Mohr J.~J., Carlstrom J.~E., Evrard A.~E., Leitch E.~M., 
2000, ApJ, 544, 629 


\bibitem []{}Hicks, A.K., \& Mushotzky, R. 2005, ApJ, 635, L9
\bibitem[]{} Leach et al.,  2008, A\&A, 491, 597
\bibitem []{}McCarthy, I.G., Balogh, M.L., Babul, A., Poole, G.B., Horner, D.J., 2004, ApJ, 613, 811
\bibitem []{}McCarthy, I.G., Babul, A., Bower, R.G., Balogh, M.L., 2008, MNRAS, 386, 1309
\bibitem []{}McNamara B. R., Rafferty D. A., Birzan L., Steiner J., Wise M. W., Nulsen P. E. J., Carilli C. L., Ryan R., Sharma M., 2006, ApJ 648, 164
\bibitem[\protect\citeauthoryear{Majumdar 
\& Mohr}{2004}]{2004ApJ...613...41M} Majumdar S., Mohr J.~J., 2004, ApJ, 613, 41 

\bibitem[\protect\citeauthoryear{Martin et al.}{2005}]{2005ApJ...619L...1M} 
Martin D.~C., et al., 2005, ApJ, 619, L1

\bibitem[\protect\citeauthoryear{Melin, Bartlett, 
\& Delabrouille}{2006}]{2006A&A...459..341M} Melin J.-B., Bartlett J.~G., Delabrouille J., 2006, A\&A, 459, 341 

\bibitem[\protect\citeauthoryear{Mittal et 
al.}{2009}]{2009A&A...501..835M} Mittal R., Hudson D.~S., Reiprich T.~H., Clarke T., 2009, A\&A, 501, 835 

\bibitem[\protect\citeauthoryear{Nagai}{2006}]{2006ApJ...650..538N} Nagai 
D., 2006, ApJ, 650, 538 

\bibitem[\protect\citeauthoryear{Nagai, Kravtsov, 
\& Vikhlinin}{2007}]{2007ApJ...668....1N} Nagai D., Kravtsov A.~V., Vikhlinin A., 2007, ApJ, 668, 1 


\bibitem[\protect\citeauthoryear{Morandi, Ettori, 
\& Moscardini}{2007}]{2007MNRAS.379..518M} Morandi A., Ettori S., Moscardini L., 2007, MNRAS, 379, 518 





\bibitem []{}O'Dea et al 2008, ApJ, 681, 1035
\bibitem[\protect\citeauthoryear{Pierpaoli et 
al.}{2005}]{2005MNRAS.359..261P} Pierpaoli E., Anthoine S., Huffenberger 
K., Daubechies I., 2005, MNRAS, 359, 261 

\bibitem []{}Pipino,  A.,  Matteucci,  F. 2004, MNRAS, 347, 968

\bibitem[\protect\citeauthoryear{Pipino et 
al.}{2009}]{2009A&A...505.1075P} Pipino A., Devriendt J.~E.~G., Thomas D., Silk J., Kaviraj S., 2009b, A\&A, 505, 1075 

\bibitem[\protect\citeauthoryear{Pipino et al.}{2009}]{2009MNRAS.395..462P} 
Pipino A., Kaviraj S., Bildfell C., Babul A., Hoekstra H., Silk J., 2009a, 
MNRAS, 395, 462

\bibitem[\protect\citeauthoryear{Peres et al.}{1998}]{1998MNRAS.298..416P} 
Peres C.~B., Fabian A.~C., Edge A.~C., Allen S.~W., Johnstone R.~M., White 
D.~A., 1998, MNRAS, 298, 416 

\bibitem[\protect\citeauthoryear{Pipino et al.}{2009}]{2009MNRAS.395..462P} 
Pipino A., Kaviraj S., Bildfell C., Babul A., Hoekstra H., Silk J., 2009, 
MNRAS, 395, 462

\bibitem[\protect\citeauthoryear{Pires et 
al.}{2006}]{2006A&A...455..741P} Pires S., Juin J.~B., Yvon D., Moudden Y., Anthoine S., Pierpaoli E., 2006, A\&A, 455, 741 


\bibitem[\protect\citeauthoryear{Poggianti et 
al.}{1999}]{1999ApJ...518..576P} Poggianti B.~M., Smail I., Dressler A., 
Couch W.~J., Barger A.~J., Butcher H., Ellis R.~S., Oemler A.~J., 1999, 
ApJ, 518, 576 


\bibitem []{}Poole G. B., Fardal M. A., Babul A., McCarthy I. G., Quinn T., Wadsley J., 2006, MNRAS, 373, 881

\bibitem[\protect\citeauthoryear{Popesso et 
al.}{2007}]{2007A&A...461..397P} Popesso P., Biviano A., B{\"o}hringer H., Romaniello M., 2007, A\&A, 461, 397

\bibitem []{}Rafferty, D. A.; McNamara, B. R.; Nulsen, P. E. J.2008, ApJ, 687, 899

\bibitem[\protect\citeauthoryear{Rasmussen et 
al.}{2006}]{2006MNRAS.373..653R} Rasmussen J., Ponman T.~J., Mulchaey 
J.~S., Miles T.~A., Raychaudhury S., 2006, MNRAS, 373, 653 


\bibitem[\protect\citeauthoryear{Reiprich {\ 
B&ouml}hringer}{2002}]{2002ApJ...567..716R} Reiprich T.~H., B{\"o}hringer H., 2002, ApJ, 567, 716 

\bibitem[\protect\citeauthoryear{Richard et 
al.}{2009}]{2009arXiv0911.3302R} Richard J., et al., 2009, arXiv, 
arXiv:0911.3302

\bibitem[\protect\citeauthoryear{Ruhl et al.}{2004}]{2004SPIE.5498...11R} 
Ruhl J., et al., 2004, SPIE, 5498, 11 



\bibitem[\protect\citeauthoryear{Sanderson, Ponman, 
\& O'Sullivan}{2006}]{2006MNRAS.372.1496S} Sanderson A.~J.~R., Ponman T.~J., O'Sullivan E., 2006, MNRAS, 372, 1496 

\bibitem[\protect\citeauthoryear{Sanderson, Edge, 
\& Smith}{2009}]{2009arXiv0906.1808S} Sanderson A.~J.~R., Edge A.~C., Smith G.~P., 2009, arXiv, arXiv:0906.1808 



\bibitem[\protect\citeauthoryear{Santos et 
al.}{2008}]{2008A&A...483...35S} Santos J.~S., Rosati P., Tozzi P., B{\"o}hringer H., Ettori S., Bignamini A., 2008, A\&A, 483, 35 


\bibitem[\protect\citeauthoryear{Sch{\"a}fer et 
al.}{2006}]{2006MNRAS.370.1713S} Sch{\"a}fer B.~M., Pfrommer C., Hell 
R.~M., Bartelmann M., 2006, MNRAS, 370, 1713 

\bibitem[]{} Sch{\"a}fer B.~M., Bartelmann M., 2007, MNRAS, 377, 253 

\bibitem[\protect\citeauthoryear{Sunyaev 
\& Zeldovich}{1970}]{1970CoASP...2...66S} Sunyaev R.~A., Zeldovich Y.~B., 1970, CoASP, 2, 66 

\bibitem []{}Stanford., S.A., Eisenhardt, P.R., $\&$ Dickinson, M. 1998, ApJ, 492, 461
\bibitem []{}Staniszewski, Z, et al., arXiv:0810.1578

\bibitem[\protect\citeauthoryear{Tran et al.}{2005}]{2005ApJ...619..134T} 
Tran K.-V.~H., van Dokkum P., Illingworth G.~D., Kelson D., Gonzalez A., 
Franx M., 2005, ApJ, 619, 134
\bibitem[\protect\citeauthoryear{Vale 
\& White}{2006}]{2006NewA...11..207V} Vale C., White M., 2006, NewA, 11, 207 

\bibitem[\protect\citeauthoryear{van Dokkum et 
al.}{2000}]{2000ApJ...541...95V} van Dokkum P.~G., Franx M., Fabricant D., 
Illingworth G.~D., Kelson D.~D., 2000, ApJ, 541, 95 


\bibitem[\protect\citeauthoryear{Vikhlinin et 
al.}{2006}]{2006ApJ...640..691V} Vikhlinin A., Kravtsov A., Forman W., 
Jones C., Markevitch M., Murray S.~S., Van Speybroeck L., 2006, ApJ, 640, 
691 

\bibitem[\protect\citeauthoryear{Vikhlinin et 
al.}{2009}]{2009ApJ...692.1033V} Vikhlinin A., et al., 2009, ApJ, 692, 1033 


\bibitem[\protect\citeauthoryear{Vikhlinin et 
al.}{2007}]{2007hvcg.conf...48V} Vikhlinin A., Burenin R., Forman W.~R., 
Jones C., Hornstrup A., Murray S.~S., Quintana H., 2007, hvcg.conf, 48 

\bibitem[\protect\citeauthoryear{Wilman et al.}{2009}]{2009ApJ...692..298W} 
Wilman D.~J., Oemler A., Mulchaey J.~S., McGee S.~L., Balogh M.~L., Bower 
R.~G., 2009, ApJ, 692, 298 


\bibitem []{}Weller, J.; Battye, R. A.2003NewAR..47..775

































\end{thebibliography}
\end{document}